\documentstyle[epsfig]{mn}
\def\figdir{.}

\def\epsscale#1{\epsfxsize=#1\columnwidth}
\def\plotone#1{\par\centerline{\epsfbox{#1}}}

\def\etal{{ et al.~\/}}

\title{Measuring Shapes of Galaxy Images II:  Morphology 
of 2MASS Galaxies}

\author[Rahman \& Shandarin]
{Nurur Rahman and Sergei F.\ Shandarin\\
Department of Physics and Astronomy,
University of Kansas, Lawrence, KS 66045, USA;\\
nurur@kusmos.phsx.ukans.edu, sergei@ku.edu}
\date{}

\begin{document}
\maketitle

\begin{abstract}
We study a sample of 112 galaxies of various Hubble types imaged in the 
Two Micron All Sky Survey (2MASS) in the Near-Infra Red (NIR; 1-2 $\mu$m) 
$J$, $H$, and $K_s$ bands. The sample contains (optically classified) 32 
elliptical, 16 lenticulars, and 64 spirals acquired from the 2MASS 
Extended Source Catalogue. We use a set of non-parametric shape measures 
constructed from the Minkowski Functionals (MFs) for galaxy shape analysis. 

We use ellipticity ($\epsilon$) and orientation angle ($\Phi$) as shape 
diagnostics. With these parameters as functions of area within the 
isophotal contour, we note that the NIR elliptical galaxies with 
$\epsilon > 0.2$ show a trend of being centrally spherical and increasingly 
flattened towards the edge, a trend similar to images in optical wavelengths. 
The highly flattened elliptical galaxies show strong change in ellipticity 
between the center and the edge. 
The lenticular galaxies show morphological properties resembling either 
ellipticals or disk galaxies. Our analysis shows that almost half of the 
spiral galaxies appear to have bar like features while the rest are likely 
to be non-barred. Our results also indicate that almost one-third of 
spiral galaxies have optically hidden bars.

The isophotal twist noted in the orientations of elliptical galaxies 
decreases with the flattening of these galaxies indicating that twist 
and flattening are also anti-correlated in the NIR, as found in optical 
wavelengths. The orientations of NIR lenticular and spiral galaxies show 
a wide range of twists.
\end{abstract}
\section{Introduction}
Galaxy morphology in different wave-bands provides useful information 
on the nature of galaxy evolution as well as the overall distribution 
of galaxy constituents such as old red giants, young luminous stars, 
gas, dust etc. For example, the younger Population I stars 
associated with massive gas-rich star formation regions light up the 
disk galaxies in optical wavelengths. The distribution of older 
Population II stars, the dominant matter component near the central 
regions of galaxies, remains hidden. The presence of interstellar 
dust hides the old stellar population especially in late-type disk 
galaxies. The NIR light, on the other hand, is much less affected by 
the interstellar dust and more sensitive to the older populations. 
Thus it provides a penetrating view of the core regions in disk 
galaxies. 
Therefore careful analysis of morphological differences between the 
optical and infrared images would not only provide valuable insight 
into the role of population classes in morphology but also reveal 
whether the discrepancies are due to singular or combined effects of 
extinction and population differences (Jarrett et al. 2003).

In the morphological studies of cosmological objects the most widely 
used technique is the ellipse-fitting method, (Carter 1978; Williams 
\& Schwarzschild 1979; Leach 1981; Lauer 1985; Jedrzejewski 1987; 
Fasano \& Bonoli 1989; Franx, Illingworth \& Heckman 1989; Peletier 
et al. 1990). 
In this study we use a set of measures known as the Minkowski 
functionals (hereafter MFs, Minkowski 1903) to analyze the morphology 
of NIR galaxies.
Contrary to the conventional method, the MFs provide a non-parametric 
description of the images implying that no prior assumptions are made 
about the shapes of the images. 
The analyses based on the MFs appear to be robust and numerically 
efficient when applied to various cosmological studies, e. g., 
galaxies, galaxy-clusters, CMB maps etc. 
(Mecke, Buchert \& Wagner 1994; Schmalzing \& Buchert 1997; Kerscher 
et al. 1997; Schmalzing \& Gorsky 1998; Hobson, Jones \& Lasenby 1999; 
Novikov, Feldman \& Shandarin 1999; Schmalzing et al. 1999; Beisbart 
2000; Kerscher et al. 2000a; Novikov, Schmalzing, \& Mukhanov 2000; 
Beisbart, Buchert \& Wagner 2001a; Beisbart, Valdarnini \& Buchert 
2001b; Kerscher et al. 2001b; Shandarin 2002; Shandarin et al. 2002; 
Sheth et al. 2003; Rahman \& Shandarin 2003, hereafter paper 1) 

This is the second in a series of papers aimed to study the morphology 
of galaxy images using a set of measures derived from the MFs. In this 
paper we analyze a larger sample of 2MASS galaxies imaged at $J, H,$ 
and $K_s$ band in NIR (Jarrett 2000; Jarrett et al. 2000; 
Jarrett et al. 2003). We have described and tested the set of Minkowski 
parameters derived from the two-dimensional scalar, vector and several 
tensor MFs to quantify galaxy shapes for a small sample of 2MASS images 
in paper I. 
The analyses in paper I used contour smoothing to reduce the effect 
of background noise. We have used the same technique in the present 
sample which contains NIR galaxies over the entire range of Hubble 
types including ellipticals, lenticulars and spirals. 
The present investigation is aimed at obtaining structural information 
on 2MASS galaxies by measuring their shapes quantified by ellipticity 
and orientation. As dusty regions of galaxies become transparent in 
the NIR, the imaging in this part of the spectrum should provide a clear 
view of the central core/bulge regions of these objects. A systematic 
study of NIR images should provide valuable information regarding the 
central structures of galaxies (e. g., optically hidden bar) which would 
otherwise remain absent when viewed in optical wavelengths. 
If only the old red giants illuminate galaxies at NIR wavelengths and 
are decoupled from Population I star lights, then the NIR galaxies should 
show weak isophotal twist in their orientations compared to those in the 
visual wavelengths. Therefore it would be interesting to check whether 
or not isophotal twist is a wavelength dependent effect.

The organization of the paper is as follows: the 2MASS sample and selection 
criteria are described in $\S2$, a brief discussion of the parameters is 
given in $\S3$. We discuss the robustness of the measures to identify and 
discern galaxy isophotes of various shapes and present our results $\S4$. 
We summarize our conclusions in $\S5$. In the appendix ($\S6$) we demonstrate 
the sensitivity of several Minkowski measures to image contamination by 
foreground stars. 
\section{2MASS Data}
The 2MASS catalogue contains near-infrared images of nearby galaxies within 
redshift range from $cz \sim 10,000$ km s$^{-1}$ to $30,000$ km s$^{-1}$. 
The survey utilizes the NIR band windows of 
$J (1.11-1.36 \ \mu m)$, $H (1.50-1.80 \ \mu m)$ and 
$K_s (2.00-2.32 \ \mu m$). The 2MASS images have 1$^{''}$ pixel resolution 
and $2^{''}$ beam resolution. The seeing FWHM values for these images are 
typically between $2.5^{''}$ and $3^{''}$ in all three bands. For details 
of the 2MASS observations, data reduction and analysis, readers are referred 
to Jarrett (2000) and Jarrett et al. (2000, 2003). 

Our sample contains 112 galaxies imaged in NIR $J$, $H$, and $K_s$ bands. 
It includes 32 elliptical, 16 lenticulars, and 64 spirals acquired from 
the 2MASS Extended Source Catalogue (XSC; Jarrett et al. 2003). The spiral 
sample contains 19 normal (SA), 21 transitional (SAB), and 24 barred (SB) 
galaxies. The galaxy types  are taken from the RC3 catalogue (de Vaucouleurs 
et al. 1992). The sensitivity and resolution ($\sim 2^{''}$) of the NIR data 
obtained in the 2MASS is not adequate to derive independent galaxy 
sub-classification (Jarrett 2000), therefore, we rely on the morphological 
classification based upon optical data derived in combination with 
both imaging and spectroscopy.

We construct the sample by hand with a moderately large number of galaxies 
of each type to make statistical inferences. The primary motivation behind 
constructing the sample is to make a comparative analysis with previous 
results and to investigate the galaxies with new tools to gain further 
insight into overall galaxy morphology in infrared wave-bands. We consider 
only bright galaxies in three different bands; the $K_s$ band total magnitude 
for the galaxies is $7 \leq K_s \leq 12$. Spiral galaxies with inclination 
up to $i \sim 60^{o}$ are included in the sample. No deprojection has been 
made to any of these galaxies prior to the analysis since the projection 
effect does not pose a serious threat to the reliability of the analysis 
when using a parameter such as ellipticity in the structural analysis of 
low inclination spiral galaxies (Martin 1995; Abraham et al. 1999). 

All galaxies in our sample are flat fielded and background subtracted. 
Except for three ellipticals, foreground stars have been removed from 
the rest of the sample. Those galaxies where a foreground star is left 
embedded in images are included purposely to illustrate the sensitivity 
of the morphological measures, as explained in the appendix.    
\section{Minkowski Functionals as Shape Descriptors}
For an object with arbitrary shape a complete morphological description 
requires both topological and geometrical characteristics. The MFs consist 
of a set of measures carrying both geometric (e. g., area, perimeter) and 
topological (the Euler Characteristic, EC) information about an object. The 
functionals obey a set of properties such as motion invariance, additivity 
and continuity (see Schmalzing 1999; Beisbart 2000). For this study we 
derive morphological parameters using a selection of two-dimensional scalar, 
vector, and tensor MFs as described in paper I. 
 
We treat every image as a set of contour lines corresponding to a set of 
surface brightness levels. A contour is constructed by linear interpolation 
at a given level. For every contour, the first step of the functional 
analysis provides three scalars: $A_S$, $P_S$, and EC (also represented by 
the symbol, $\chi$); three vectors or centroids: $A_i$, $P_i$, and $\chi_i$; 
and a total of nine components of three symmetric tensors $A_{ij}$, $P_{ij}$, 
and $\chi_{ij}$. Here $i, j = 1, 2$ (for details see paper 1). In the next 
step, the eigenvalues ($\lambda_1$ and $\lambda_2$; $\lambda_1 > \lambda_2$) 
of the tensors are found taking centroids as the origins of corresponding 
tensors. 
After calculating the eigenvalues, we proceed to construct the axes and 
orientations of the ``auxiliary ellipse'' (hereafter AE). To construct the 
area tensor AE, for example, we take the eigenvalues of $A_{ij}$ and ask 
what possible ellipse may have exactly the same tensor. When we find that 
particular ellipse, we label it as the area tensor AE. The orientation of 
the semi-major axis of the AE with respect to the positive x-axis is taken 
as its orientation.  
The AEs corresponding to the perimeter and EC tensors are constructed in 
a similar manner. To discern morphologically different objects, therefore, 
we use ellipticities ($\epsilon_i$) and orientations ($\Phi_i$) of the AEs 
rather than the eigenvalues of the tensors. We define ellipticity of the 
AEs as $$\epsilon_i = 1 - b_i/a_i,$$ where $i$ corresponds to one of the 
three tensors, and $a$ and $b$ are the semi-major and semi-minor axes of 
the AEs.

The use of AEs effectively relates a contour to an ellipse: the similarity 
of three AEs is a strong evidence that the shape of the contour is 
elliptical. For example, in case of a perfect elliptic contour, the AEs 
will be the same. In particular, the areas of all three AEs will be equal 
to the area of the contour, i. e., $A_A = A_P = A_{\chi} = A_S$, and  
the perimeters of the ellipses will be equal to the perimeter of the 
contour, i. e., $P_A = P_P = P_{\chi} = P_S$. 
In addition, the orientations of all three ellipses will coincide with 
the orientation of the contour. Therefore, if plotted, all three AEs will 
be on top of each other, overlapping with the contour. For that contour, 
all three vector centroids will also coincide with each other and with the 
center of the contour. 
Note that the latter alone does not guarantee that the contour itself 
is elliptical in nature since for any centrally symmetric contour 
the centroids would coincide. However, for a non-elliptical contour all 
three AEs will be different in size and orientation (see Fig. 1, 
paper I). 

\begin{figure}
\epsscale{1.1}
\plotone{\figdir/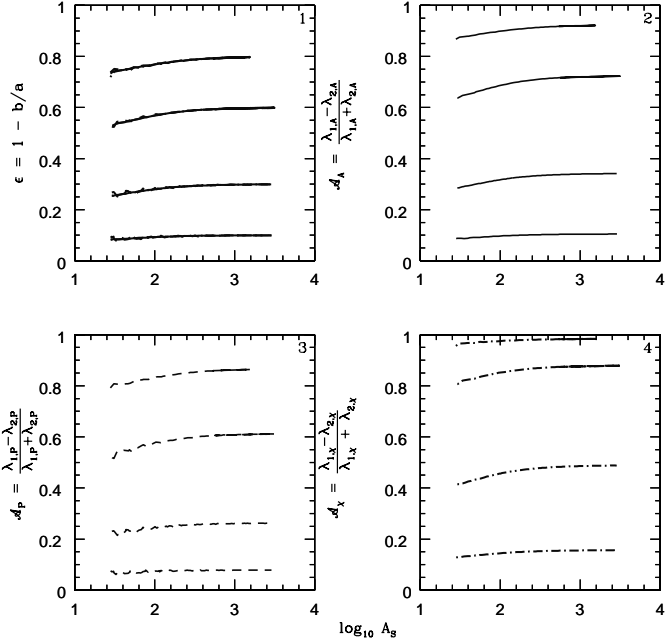}
\caption{Ellipticity ($\epsilon_i$) and Anisotropy (${\cal A}_i$) as a 
function of contour area ($A_S$) for various elliptic profiles. For each 
profile, panel 1 have $\epsilon_i$ of three AEs whereas panels 2, 3, and 
4 show, respectively, ${\cal A}_A$, ${\cal A}_P$, and ${\cal A}_{\chi}$. 
In the figure solid, dashed and dashed-dotted lines represent, respectively, 
the parameters derived from the area, perimeter, and EC tensor. 
For perfect elliptic profiles, the lines showing $\epsilon_i$ stay on top of 
each other giving the identical result. The AEs behave similarly for elliptic 
profiles of arbitrary  flattening. 
The ${\cal A}_i$, however, do not coincide. The relative separations between 
the lines representing different tensors change with the flattening of the 
profiles. Note that for highly flattened systems central regions appear less 
elongated due to discreteness effect. 
\label{scaling}}
\end{figure}

Note that the sets of eigenvalues from three tensors can be used to 
construct three ``anisotropy'' parameters instead of three AEs (see 
also Beisbart 2000). The parameters can be defined as 
$${\cal A}_i = \frac{\lambda_{1,i} - \lambda_{2,i}}
{\lambda_{1,i} + \lambda_{2,i}},$$ where $i$ has the same meaning as before. 
To better understand the behaviors of $\epsilon_i$ and ${\cal A}_i$, we 
show these parameters in Fig. \ref{scaling} as functions of contour area 
($A_S$) for four elliptic profiles of different flattening. Panel 1 of the 
figure shows ellipticity for all four profiles. Panels 2, 3, and 4 show, 
respectively, ${\cal A}_A$, ${\cal A}_P$, and ${\cal A}_{\chi}$ where each 
of these panels also has four profiles.      
For each profile, we find that the $\epsilon_i$ of the AEs are 
identical and coincide with each other (panel 1). The ${\cal A}_i$, on the 
other hand, do not coincide even for perfect elliptic contours (panels 2, 3, 
and 4). We also find that the relative separations between different 
${\cal A}_i$ change with the flattening of the contours. From the behavior 
of the parameters one can think that the $\epsilon_i$ of the AEs act as 
parameters that are scaled with respect to the ${\cal A}_i$. For contours 
with arbitrary flattening, the ${\cal A}_i$ from the area and EC tensors 
need to scale down to match with the $\epsilon_i$ of the respective AEs. 
The ${\cal A}$ from the perimeter tensor, however, needs to scale up for 
spherical and moderately elongated contours. For highly elongated contours, 
however, all three ${\cal A}$s need to scale down. 

The illustration of the ``anisotropy'' parameter serves two purposes. 
First, it gives us a feeling of the AEs compared to the conventional 
parameter that deals with eigenvalues. Second, it demonstrates that one 
can derive various shape measures from the set of MFs. Apart from this 
parameter, one can also derive the shapefinder statistic as suggested by 
Sahni, Sathyaprakash \& Shandarin (1998). However, we restrict ourselves 
to ellipticity and orientation since these are the two widely used measures 
in astronomy. We will explore the sensitivity and robustness of ${\cal A}_i$ 
in our future work on optical galaxies. 

The non-parametric approach for shape analysis, such as moments technique, 
has been known to the astronomical community for some time (Carter 1978; 
Carter \& Metcalf 1980). It should be mentioned here that the morphological 
analyses based only on the moments of inertia would provide incomplete and 
sometimes misleading results. 
As an example, let us assume that one has a galaxy image which has been 
kept in a black box and analyzed using simply the inertia tensor without 
having a priori knowledge of the shape of the image. The analysis based 
only on the moments of inertia will provide a resultant ellipticity of the 
object regardless of its actual shape. Using this result one can always 
infer an elliptical shape for the unseen object. If one raises the question 
of the likeliness of the elliptical shape of the object, the analysis based 
on the inertia tensor alone will not be able to give a satisfactory answer. 
One needs to invoke additional measure(s) in order to justify the result. It 
is at this point where the measures derived from the set of MFs appear to be 
effective. Subsequent analyses of the image using moments of the perimeter 
and EC tensor enables one to pin down the type of the galaxy and thus ensures 
the objectivity of the analysis. 

The ellipticities obtained from different AEs provide information (regarding 
shapes) similar to the conventional shape measure based on  inertia tensor. 
The main difference is that the conventional method finds the eigenvalue 
of the inertia tensor for an annular region enclosing mass density or 
surface brightness. The method based on MFs, however, finds the eigenvalues 
of contour(s) where the region enclosed by the contour(s) is assumed to be 
homogeneous and to have constant surface density.   

In order to reduce the effect of noise present in the image we use a simple 
smoothing technique. Instead of smoothing the whole image, we smooth contours 
at each brightness level using the procedure known as the unequally weighted 
moving average method. The goal of this smoothing is to restore the initial 
unperturbed contour as much as possible and measure its morphological 
properties. The implementation of the smoothing is described in detail in 
paper I. 

Contour smoothing considerably improves the estimates of ellipticity derived 
from the tensor functionals (see Figs. 8 and 9, paper I), therefore, in 
this study, we focus on the results obtained only from the smoothed contours. 
However, we note two effects that arise as a result of smoothing. First, in 
the outer regions of galaxies a smoothed, outer contour crosses the inner one. 
It happens occasionally; however, the area within the smoothed, outer contour 
remains greater than the inner one and does not affect the measurement. 
Second, in case of very large contour, we loose information. The smoothing 
technique is an iterative process depending on the number of points along a 
contour (see paper 1 for details). A highly irregular contour that consists 
of large number of points eventually shrinks to a point due to excessive 
smoothing. As a result we do not get any contribution from it. 
However, this does not pose any threat on getting relevant information of the 
shape of image contours since we can always compare with the unsmoothed 
profiles to see how much information is lost.  
\section{Results}
In this paper we use a different notation to express ellipticity than in 
paper I where we used $E = 10*(1 - b/a)$ in the range 0 to 10. The symbol 
$E$ for the ellipticity parameter is often used for the definition of the 
Hubble types for ellipticals and this symbol and range can be misleading 
when used as a parameter for other types of galaxies (e. g., spirals). 
We, therefore, use $\epsilon_i$ as the characteristic of shape with the 
range 0 to 1.

The ellipticity and orientation can be used to isolate galaxies into two 
broad bins: non-barred and barred systems. For finer distinction we need 
to use other parameters such as luminosity, color, surface brightness, 
half-light radius, etc. Both of these parameters have been applied to 
quantify properties of elliptical galaxies (e.g. Carter 1978; Williams 
\& Schwarzschild 1979; Leach 1981, Lauer 1985; Jedrzejewski 1987; Fasano 
\& Bonoli 1989; Franx, Illingworth \& Heckman 1989; Peletier et al. 1990). 
In many recent works this set of parameters has been used to obtain 
information on different structural components, e. g., bar, bulge, disk 
etc. of disk galaxies (Athanassoula et al. 1990; Martin 1995; Wozniak et 
al. 1995; Rozas, Knapen \& Beckman 1998; Abraham \& Merrifield 2000; 
Laurikainen, Salo \& Rautainen 2002; Peng et al. 2002, Erwin \& Sparke 
1999; Erwin \& Sparke 2003; Michel-Dansac \& Wozniak 2004).

Galaxies with different morphologies appear to show characteristic signatures 
in the ellipticity profile. For example, the profile of elliptical galaxies 
is generally monotonic. A barred galaxy, however, shows a distinct peak with 
a spherical central region, signaling the presence of a central bulge and a 
bar. On the other hand, for a multi-barred system several peaks appear in the 
profile (Wozniak et al. 1995; Erwin \& Sparke 1999). 

Elliptical galaxies usually show twists (i. e., a change in the orientation 
of AE) in the orientation profiles with varied strength depending on the 
flattening of the contours. Barred galaxies, on the other hand, not only 
have large twists in their orientations but also have characteristic 
features such as a sharp peak, or two different but approximately constant 
orientations with an abrupt change in between (Wozniak et al. 1995; Erwin 
\& Sparke 1999). 

In the contemporary studies, therefore, a barred galaxy is identified as 
the system whose ellipticity and orientation profile simultaneously show 
the distinctive signatures mention above. In this study we follow the same 
criterion to analyze disk galaxies, i. e., we identify bars by visual 
inspection with the condition that both ellipticity and orientation profiles 
show the characteristic features simultaneously. However, identifying disk 
galaxies as barred systems with the above criterion (as used in previous 
studies mostly in optical wavelengths) should be used with caution. 
Our experience shows that optically classified barred galaxies appear with 
distinct peaks in their ellipticity profiles but show continuous orientation 
over the region where the peak in the shape profiles persists. Therefore we 
feel that more elaborate treatment is needed for identifying barred systems 
rather than simply relying on the behaviors of ellipticity and orientation. 
In this study, therefore, when we encounter these types of systems, we 
refrain from drawing any conclusions about these galaxies. We will explore 
the detailed shape properties of these galaxies in our future study including 
other structural measures such as Fourier decomposition technique (Quillen, 
Frogel \& Gonzalez 1994; Buta \& Block 2001).  
  
We measure ellipticity and orientation for a set of contours obtained at 
different surface brightness levels. Generally these two parameters vary 
with the area of the contour, i. e., 
$\epsilon_i = \epsilon_i (A_S)$ and $\Phi_i = \Phi_i (A_S)$. 

We analyze each galaxy at 30 different brightness levels where the levels 
correspond to equally spaced areas on a logarithmic scale covering almost 
the entire region of each galaxy. At every brightness level, contours are 
found and subsequently smoothed. All three AEs are constructed from the 
smoothed contours. In order to reduce information content we present our 
final results using only the area tensor AE and hence drop the subscript 
$i$. Below we briefly discuss the justification of the choice. For each 
galaxy, therefore, we show $\epsilon$ and $\Phi$ of this particular AE as 
a function of contour area ($A_S$).

For each galaxy, a thick dashed line is used to show the mean values of 
$\epsilon$ and $\Phi$ calculated from $J, H,$ and $K_s$ bands. The mean 
values are used to estimate the overall change ($\Delta \epsilon$ and 
$\Delta \Phi$) which is defined as the difference between the highest 
and lowest value of the corresponding mean value. It is a single number 
and independent of $A_S$.  
Two thin solid lines are used to show the maximum and minimum of the 
measures obtained from different bands. The difference between these 
two thin solid lines is used to quantify the scatter ($\delta \epsilon$ 
and $\delta \Phi$) in the parameters. The $\delta$ varies at different 
regions of a galaxy and so depends on $A_S$. We use it as an indicator 
to measure the dependence of galaxy shapes on different colors.

For each sample of galaxies, the result is presented in increasing order 
of 2MASS ellipticity obtained from the 2MASS catalogue. The 2MASS 
ellipticity and orientation are measured for the $3 \times \sigma_n$ 
isophote in the $K_s$ band ($\sigma_n$ is the rms amplitude of the 
background noise provided by the catalogue). 
For each galaxy, the 2MASS shape parameter is, therefore, a single number. 
Our analysis, on the other hand, provides a range of values obtained at 
different regions (recall that $\epsilon$ or $\Phi$ is a function of area) 
from the galactic center. We use the 2MASS estimate of ellipticity and 
position angle as the reference. It is shown by the horizontal dashed line. 
In all plots the vertical dashed line represents the contour area ($A_S$) 
corresponding to $K_s$ band $3 \times \sigma_n$ isophote. We rescale the 
orientation profiles of a few galaxies to fit the desired ranges that have 
been chosen to show the $\Phi$ profiles.

Note that our results presented below correspond to the area 
$\log_{10} A_S \geq 1.5$. The range is chosen in order to exclude 
discreteness effects due to the grid. The deviation ($\Delta$) as well as 
the scatter ($\delta$) in the parameters, therefore, will be considered in 
this range.

\subsection{Shape of Isophotal Contour and The Role of Tensor Ellipses}
From section $\S 3$ we know that for a perfect elliptic contour, the 
areas enclosed by the AEs will be the same  whereas for non-elliptic 
contours they will be different. Therefore, comparing the areas enclosed 
by different ellipses one can probe to which extent the shape of a 
contour can be approximated by an ellipse. 

\begin{figure}
\epsscale{1.1}
\plotone{\figdir/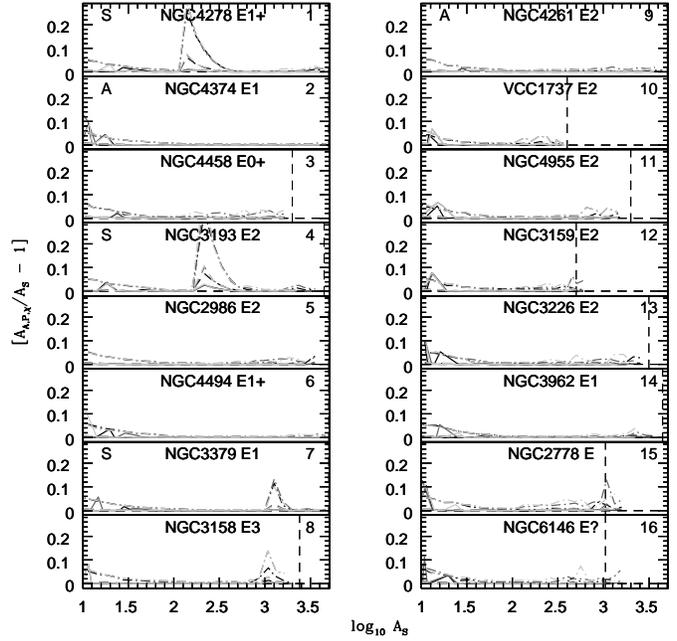}
\caption{The relative difference in the areas enclosed by three different 
AEs as a function of contour area ($A_S$) for a selection elliptical 
galaxies. 
Each panel contains a total of nine curves : three curves corresponding to 
three AEs at each band. The dark, medium and light colors to represent $J$, 
$H$, and $K_s$ band, respectively. 
The area, perimeter, and EC ellipses are shown, respectively, by the solid, 
dotted, and dashed-dotted lines. The vertical dashed line represents the 
area within the contour that corresponds to $K_s$ band $3 \times \sigma_n$ 
isophotal region. For some galaxies this line goes beyond the horizontal 
range shown above. The number 1 to 16 shown at top-right in each panel is 
used just to label the galaxies. \label{aa_1}}
\end{figure}

\begin{figure}
\epsscale{1.1}
\plotone{\figdir/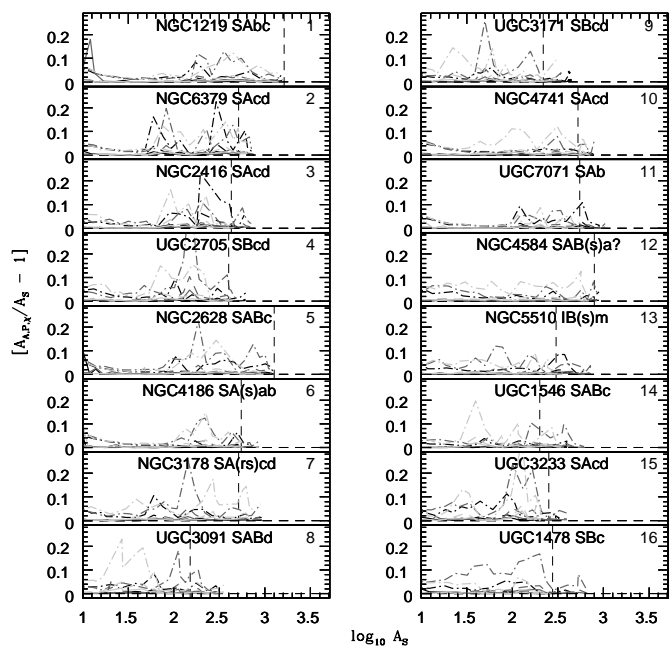}
\caption{The relative difference in the areas enclosed by different AEs as 
a function of contour area ($A_S$) for a selection of spiral galaxies. 
The line styles and colors are similar to Fig. \ref{aa_1}. \label{aa_2}}
\end{figure}

Galaxies appear more regular in NIR wave bands than in the visual 
wavelengths. For example, a galaxy may appear grand in the visual bands 
with its giant spiral arms. However, it will lose much of its grandeur 
in infrared wave-bands. Since spiral arms consist mostly of gas and 
young bright stars, they will be absent in long wavelength parts of the 
spectrum. 
To see up to what extent galaxy contours of different Hubble types retain 
their characteristic signatures in NIR and how tensor ellipses help us to 
understand their shapes, we draw readers attention to Figs. \ref{aa_1} 
and \ref{aa_2}. In these figures we show the relative difference in areas 
enclosed by three different AEs as a function of contour area ($A_S$) for 
a selection of elliptical and spiral galaxies. 
The figures highlight only the interesting parts along the vertical axis. 
Each panel contains a total of nine curves: three curves corresponding to 
three AEs from each band. The dark, medium and light colors represent, 
respectively, the $J$, $H$, and $K_s$ band.
We show the area, perimeter, and EC ellipses, respectively, by the solid, 
dotted, and dashed-dotted lines.

Apart from three galaxies that are marked by ``S'', the isophotal contours 
of most of the galaxies in Fig. \ref{aa_1} show elliptic nature in the NIR. 
Few of these, e. g., NGC 3158 and NGC 2778 (galaxy number 8 and 15 
respectively) appear slightly non-elliptic in one or two bands. The galaxies 
marked by ``S'' (number 1, 4, and 7) are galaxies which have foreground 
stars embedded in their image. For these galaxies the notable feature is 
the sharp increase in area around the region of the images where the star 
is embedded. One can see that the same ellipse overlaps in different bands. 
This is most notable from the EC ellipse (dashed-dotted line) since the EC 
tensor is the most sensitive to any disturbance along the contour. The 
other two ellipses also behave in a similar manner but with less sensitivity. 
This particular behavior shows that the AEs detect unusual features 
attached to otherwise perfectly elliptical image body. For example, 
when we compare NGC 3158 or NGC 2778 with those three galaxies, we notice 
that $A_{\chi}$ not only differs in different bands but it also spreads 
out differently around the edge. It tells us that the contours of these 
galaxies are simply non-elliptic around the edge without any abnormal 
feature. 

When we compare galaxies of different Hubble types, we find that almost all 
spiral galaxies (Fig. \ref{aa_2}) show more non-elliptic nature than the 
elliptical galaxies. The non-elliptic nature of spiral galaxy isophotes is 
reflected strongly in the behavior of the EC ellipse in all bands. For these 
galaxies, the EC ellipse is different not only from other two ellipses but 
it is also spread out arbitrarily in different bands. 

From Figs. \ref{aa_1} and \ref{aa_2}, it is clear that all three AEs are 
useful for better characterization of a galaxy image. The information 
provided by the AEs definitely help to get finer distinction of galaxies. 
Since our current interest is to focus on the gross morphological features 
rather than looking at the finer details of galaxy isophotes, to reduce the 
information content, therefore, we will only use the area tensor AE to 
present our final results.
\subsection{Ellipticals}
The sample of elliptical galaxies has been divided into two groups. 
Group 1 contains galaxies with small ellipticities ($\epsilon \leq 0.2$) 
while Group 2 has galaxies with $\epsilon > 0.2$. 
To simply locate a galaxy within a group, we label them with an integer 
number. For example, NGC 4374(2) means that NGC 4374 is the second galaxy  
in its relevant group. In each group galaxy number 1 has the lowest 2MASS 
ellipticity whereas number 16 has the highest value. We follow the 
same format for all types of galaxies. Note that below we highlight only 
those galaxies that have interesting/unusual morphological features. The 
general trends of galaxies will be investigated in $\S5$.  

\begin{center} {\bf Group 1} \end{center}

{\bf Ellipticity:}
The ellipticity profiles of galaxies in this group are shown in Fig. 
\ref{ellpe1}. 
NGC 4374(2) and NGC 4261(9), have uncommon profiles compared to other 
galaxies. These galaxies are more elongated in the region 
$1.8 < \log_{10} A_S < 3.0$ than either around the center or near the 
edge which makes their profiles look ``centrally arched'' (marked 
by ``A''). This particular type of behavior is shown only by sphericals 
and is absent in elongated galaxies. 

The following three galaxies are shown with an ``S'' mark in Fig. 
\ref{ellpe1}: NGC 4278(1), NGC 3193(4), and NGC 3379(7). These galaxies 
have sharp kinks in the $\epsilon$ profiles. These sharp changes are 
caused by a foreground star and do not correspond to any structural 
feature (see appendix). The rest of the galaxies show profiles typical 
of ellipticals. Galaxies in this group, in general, have small scatter.

{\bf Orientation:}
The orientation profiles of galaxies are presented in Fig. \ref{ange1}. 
NGC 4458(3) is stable from the center up to $\log_{10} A_S \approx 2.8$ 
and shows a large deviation near the edge. 
NGC 3193(4) shows a notable peak in its orientation. Its profile shows a 
dip around the region where the foreground star is embedded. NGC 3379(7) 
does not show any unusual feature in its profile. Its position angle, as 
well as the scatter, gradually increases towards the galactic center. 
NGC 3379(7) has larger scatter at all distances from the galactic center 
than NGC 4278(1) and NGC 3193(4). 

Note that the presence of a star on the galaxy contours may or may not be 
detected by the $\Phi$ profile. The $\epsilon$ profile, on the other hand, 
is sensitive to any kind of disturbance on the contour. If the unusual 
feature on the contour attributed by the star happens to be along the major 
axis, which is the case for NGC 3379(7), it remains unnoticed and no peak 
appears in the $\Phi$ profile. 
This is the reason why the presence of a star on the contours may be missed 
as a signature on the $\Phi$ profile. On the other hand, if the feature is 
slightly off the major axis, it changes the orientation of the contour 
significantly with respect to the inner and  outer contours. This is the 
case for NGC 4278(1) and NGC 3193(4).  

In spite of being a spherical galaxy (E1+), NGC 4494(6) has remarkably low 
twist ($\Delta \Phi \sim 3.5^{o}$) and scatter in its orientation. It is 
very unusual because a spherical contour is highly susceptible to background 
noise. A slight perturbation caused by the noise distorts the contour's 
shape and the direction of perturbation becomes its orientation.

VCC 1737(10), NGC 3159(12), and NGC 3226(13) have orientation profiles 
quite similar to disk galaxies as shown later in this section. The $\Phi$ 
profiles of VCC 1737(10) and NGC 3226(13) appear ``arch like'' where the 
latter galaxy has a longer arch than the former. These three galaxies also 
have large scatter in their orientations.

\begin{center} {\bf Group 2} \end{center}

{\bf Ellipticity:}
The ellipticity profiles of galaxies in this group are shown in Fig. 
\ref{ellpe2}. Galaxies with $\epsilon > 0.2$ show a ``centrally spherical'' 
trend in their profiles. The variation in the profile becomes stronger with 
the increasing flattening of the galaxies. The overall change in $\epsilon$ 
observed from the center to the edge varies from galaxy to galaxy. It 
changes from as low as $\sim 0.12$ (NGC 315(2)) to as high as $\sim 0.33$ 
(NGC 5791(15)). 

The measurements for most of the galaxies in all three bands nicely coincide. 
As a result the galaxies in this group show the lowest amount of scatter.

{\bf Orientation:}
The orientation profiles of galaxies in this group are shown in Fig. 
\ref{ange2}. The flattened ellipticals have quite stable orientations with 
relatively low twist and scatter compared to those in group 1. For all 
galaxies the $\Delta \Phi$ is within $10^{o}$. 

The relatively small twist observed in the elliptical galaxies with 
$\epsilon > 0.2$ suggests that flattening and isophotal twists are most 
likely anti-correlated in the NIR wave-bands (see $\S6$ and Fig. 
\ref{galletta} for more). This result is in agreement with the (optical) 
correlation found by Galletta (1980) who noted that the maximum apparent 
flattening and the highest observed twist are inversely related.    
\subsection{Lenticulars} 
The sample of lenticulars contains 16 galaxies. The ellipticity and 
orientation of these galaxies are shown in Figs. \ref{ellps0} and 
\ref{angs0} respectively.

{\bf {Ellipticity:}}
For NGC 1315(1), our estimate indicates a spherical inner part 
($\epsilon \sim 0.05$) in between $1.2 \leq \log_{10} A_S \leq 2.6$, 
suggesting a spherical bulge for the galaxy. 
NGC 3598(7), IC 4064(8), NGC 3816(12), NGC 3768(13), and NGC 2577(16) 
have profiles quite similar to elliptical galaxies: a smoothed variation 
in elongation with radius. 

{\bf {Orientation:}}
The orientations of lenticular galaxies generally have larger scatter 
than the elliptical galaxies. NGC 4659(4) shows the largest twist in 
this sample, ($\Delta \Phi \sim 68^{o}$); its orientation experiences 
a large change  within $1.5 \leq \log_{10} A_S \leq 2.3$. 
   
Lenticular galaxies with (2MASS) ellipticity $\epsilon > 0.30$ (from 
galaxy number 8 and above in Fig. \ref{angs0}) are observed to have less 
scatter than their spherical counterparts. This property is quite similar 
to flattened ellipticals. The scatter is large for lenticulars that are 
spherical in shape and it varies at different regions from the galaxy 
centers.  

$\bullet$
Both shape and orientation profiles of NGC 4620(9) and NGC 2544(14) 
exhibit a bar like feature. These galaxies are labeled as ``IB'' where 
``IB'' stands for ``infrared bar''. 
\subsection{Spirals}
The sample of spirals has been divided into four groups based on 
the degree of scatter ($\delta \epsilon$) observed in the $\epsilon$ 
profiles. $\delta \epsilon$ is measured for all $A_S$ in the range 
$\log_{10} A_S \geq 1.5$.  
Group 1 contains spirals that have the least scatter while group 4 
has galaxies with the largest scatter. 
Galaxies with $\delta \epsilon \leq 0.05$ are included into group 1, 
for group 2 the range is $0.05< \delta \epsilon \leq 0.1$, those with 
$0.1 < \delta \epsilon \leq 0.2$ are in group 3, and finally galaxies 
with $\delta \epsilon > 0.2$ define group 4. 
The grouping of the spiral sample has been done by visual examination 
of the ellipticity profiles and therefore it is quite crude; our 
intention is simply to highlight the interesting features apparent in 
the shape profiles of spiral galaxies viewed in the NIR. In these 
groups, galaxies are organized by increasing order of (2MASS) 
ellipticity.

\begin{center} {\bf Group 1} \end{center}

{\bf {Ellipticity:}}
The ellipticity profiles of galaxies in this group are shown in 
Fig. \ref{ellps1}. Only four galaxies are optically classified (RC3) 
as barred systems. These are NGC 4262(2), NGC 4024(3), NGC 3384(10), 
and NGC 4394(13). Except the last one, the shape profiles of other 
three galaxies posses distinctive feature, i. e., a clear appearance 
of peak(s) in the profiles, indicating (qualitatively) similar 
morphology both in optical and NIR. 
These three galaxies are labeled as ``OIB'' where ``OIB'' stands for 
``optical and infrared bar''. We label NGC 4394(13) as ``OB'' where 
`OB'' stands for ``optical bar'' only. We give our explanation below 
for this labeling.

NGC 4151(6) is a highly spherical galaxy compared to other members of 
this group. It has $\epsilon < 0.05$ from the galactic center up to 
$\log_{10} A_S \approx 3.2$; beyond this limit the ellipticity increases 
a little bit. This galaxy, an intermediate ringed transitional spiral, 
is most likely dominated by a very large spherical bulge. 
The profiles of IC 3831(8), NGC 4143(9), NGC 3418(11), and NGC 5326(12) 
are quite similar to elliptical galaxies. 

NGC 3912(15) and UGC 5739(16) are two highly flattened galaxies in this 
group. Their profiles appear ``centrally arched'', an unusual feature 
shown by several NIR elliptical galaxies.

{\bf {Orientation:}}
The orientation profiles of galaxies in this group are presented in Fig. 
\ref{angs1}. The overall twists ($\Delta \Phi$) in these galaxies range 
in between $\sim 5^{o}-55^{o}$. 

NGC 4151(6), a galaxy of spherical shape, shows the largest scatter in 
its orientation. Since highly spherical contours can be perturbed easily 
by noise, it is not surprising  that the galaxy would have large variations 
in its orientation at different wave-bands. 
The profile of NGC 4143(9) is quite similar to the barred galaxy NGC 
3384(10), although its orientation changes smoothly compared to that of 
NGC 3384(10). 

$\bullet$
The profiles of NGC 3177(4) and NGC 3684(5) show a bar like feature although 
these galaxies are optically classified as non-barred systems. However, we 
label these galaxies as ``IB'' after careful visual inspection of their 
unsmoothed and smoothed profiles. 

The profiles of NGC 4651(7) and NGC 4494(13) deserve attention because of 
their complex morphologies. 
The ellipticity profile of NGC 4651(7) has a peak that is confined within 
$2.5 < \log_{10} A_S < 3.0$. In this region its orientation is different 
from the bulge and the disk. Since it is an early type spiral with $rs$ 
subclass, we think the feature is due to the ring.
According to RC3 classification NGC 4494(13) is a barred galaxy. But from 
the NIR profile it is not clear whether the galaxy really possesses a 
bar like structure. The galaxy is devoid of any distinct peak in its shape 
profile even-though it has considerable amount of twist 
($\Delta \Phi \sim 32^{o}$). It shows higher flattening near the edge 
which is probably the elongation of the disk. We are uncertain about 
the actual shape and therefore label the galaxy as ``OB'' (optical bar).

\begin{center} {\bf Group 2} \end{center}

{\bf {Ellipticity:}}
The ellipticity profiles of galaxies in this group are shown in Fig. 
\ref{ellps2}. 
It contains seven galaxies that are optically classified as barred 
systems. The galaxies are NGC 3974(1), IC 357(7), NGC 4662(10), NGC 
6347(11), NGC 5737(13), UGC 2585(14), and IC 568(16). These galaxies 
show a bar like feature in their NIR ellipticity profiles. 
We, therefore, label them as ``OIB''.
The profiles of NGC 3277(2) and NGC 4714(8) are quite similar to 
elliptical galaxies. 

{\bf {Orientation:}}
The orientation profiles of the galaxies in this group are shown in Fig. 
\ref{angs2}. The overall twists range between $\sim 5^{o}-71^{o}$. 
The profiles of NGC 3974(1) and IC 357(7) have similar characteristics: 
a constant direction along the central bars while the orientation changes 
in the central bulge and outside disk. 

%
$\bullet$
The following three galaxies are optically-classified non-barred spirals: 
NGC 4146(3), IC 863(6), and UGC 5903(9). However, these galaxies show 
bar like features in their NIR profiles and therefore we label them as 
``IB''. 

Note that the orientation profile of NGC 4152(4) has two distinct peaks. 
Its ellipticity profile, however, does not possesses any feature to be 
considered as a barred galaxy. Therefore we are not certain whether or 
not it can be considered as a barred galaxy.

The orientation profile of NGC 5478(12) shows two different directions.
Its shape changes considerably around the region where the position angle 
changes. The variation in ellipticity is extended over a wide region. 
From visual inspection a question can be raised about whether or not the 
galaxy is a barred system. A careful inspection of both unsmoothed and 
smoothed profiles indicates that it is not a barred galaxy.

\begin{center} {\bf Group 3} \end{center}

{\bf {Ellipticity:}}
The ellipticity profiles of galaxies in this group are shown in Fig. 
\ref{ellps3}. This group contains the largest number of optically classified 
barred galaxies: NGC 5716(3), UGC 3936(4), NGC 4375(6), UGC 7073(9), 
NGC 3811(11), IC 742(12), NGC 5260(13), IC 1764(15), and UGC 3839(16). Except 
for NGC 4375(6) and UGC 7073(9), the NIR profiles of other 7 galaxies show 
similarity (at least qualitatively) with their optical morphology (``OIB''). 
The profiles of both NGC 4375(6) and UGC 7073(9) lack characteristic 
signatures in their profiles and therefore have been labeled as ``OB''. 

{\bf {Orientation:}}
The orientations of galaxies are shown in Fig. \ref{angs3}. The overall 
twists range between $\sim 10^{o}-71^{o}$. The twist as well as scatter 
in the orientations of these galaxies are high compared to those in the 
previous two groups. 

Among ``OIB'' galaxies, only UGC 3936(4) and UGC 3839(16) have large twists 
in their orientations. For other ``OIB'' galaxies, the twists are relatively 
small.

%
$\bullet$
The following three galaxies are optically classified as non-barred spirals: 
UGC 2303(1), UGC 3053(2), and NGC 3618(8). The profiles of these galaxies in 
NIR, however, show features similar to barred systems. We label these galaxies 
as ``IB'' after visual inspection of their unsmoothed and smoothed profiles.

\begin{center} {\bf Group 4} \end{center}

{\bf {Ellipticity:}}
The ellipticity profiles of galaxies in this group are shown in Fig. 
\ref{ellps4}. According to RC3 catalogue, the following four 
galaxies are barred systems: UGC 2705(4), UGC 3171(9), NGC 5510(13), 
and UGC 1478(16). It is interesting to note that the NIR profiles of 
none of these galaxies have characteristic features resembling 
bar like systems. Therefore we label these galaxies as ``OB''.

{\bf {Orientation:}} 
The orientations of galaxies in spiral group 4 are shown in Fig. 
\ref{angs4}. The overall twists range between $\sim 15^{o}-70^{o}$. 
Galaxies in this group have the highest scatter in their orientations 
compared to those in the other three groups.

Among ``OB'' galaxies, UGC 1478(16) has a unique profile. The galaxy 
has a distinct, highly flattened, central bar. Its orientation is 
constant along the central bar and changes sharply to the orientation 
of the disk.

%
$\bullet$
The following three galaxies are optically-classified non-barred systems: 
NGC 1219(1), UGC 3091(8), and UGC 7071(11). 
The NIR profiles of these galaxies, however, show features similar to 
barred galaxies (``IB''). These spirals generally have larger twists 
compared to the ``OB'' galaxies.
\subsection{Comparison with the 2MASS Estimate}
We use shape parameters estimated by the 2MASS as the reference. Our analysis 
should provide estimates similar to the 2MASS for contours enclosing larger 
areas. Since $3 \times \sigma_n$ isophote corresponds to a low surface 
brightness level, i. e., a region near the edge of a galaxy image, we would 
expect agreement only in this region. 

For the purpose of comparison with we use only the relative error between 
the shape of the contour closest to $K_s$ band $3 \times \sigma_n$ isophote 
and the 2MASS estimate. In most cases we find excellent agreement (within 
$\sim 10\%$ of the 2MASS). However, in a few cases the agreement is poor 
(beyond $\sim 10\%$ of the 2MASS) and in this section we report only these 
cases. 

We want to stress that the result presented here is obtained after careful 
comparison of both unsmoothed and smoothed profiles of the parameters. 
We do not include galaxies whose unsmoothed profiles agree with the 2MASS 
(i. e., within $\sim 10\%$ of the 2MASS). For example, we do not include 
NGC 4394 (13, spiral group 2; Fig. \ref{ellps2}) in the list of galaxies 
given below since its unsmoothed profile shows excellent agreement with the 
2MASS. The apparent mismatch between the 2MASS and our estimate for smoothed 
profile is due to the effect of smoothing (as mentioned in $\S3$). 
If the unsmoothed profile of a galaxy shows disagreement (i. e., beyond 
$\sim 10\%$ of the 2MASS) in the first place, it is also reflected in its 
smoothed profile and only then we include the galaxy in the list. For these 
galaxies, therefore, the estimates of the parameters from both unsmoothed 
and smoothed profiles disagree with the 2MASS. We note that that an agreement 
in the $\epsilon$ profile a galaxy does not necessarily mean that it will 
also have an agreement in its $\Phi$ profile.

Note particularly that for several elliptical galaxies we do not have any 
contour near the $3 \times \sigma_n$ level. We do not attempt to make any 
comparison for these galaxies. 
 
\begin{figure}
\epsscale{1.15}
\plotone{\figdir/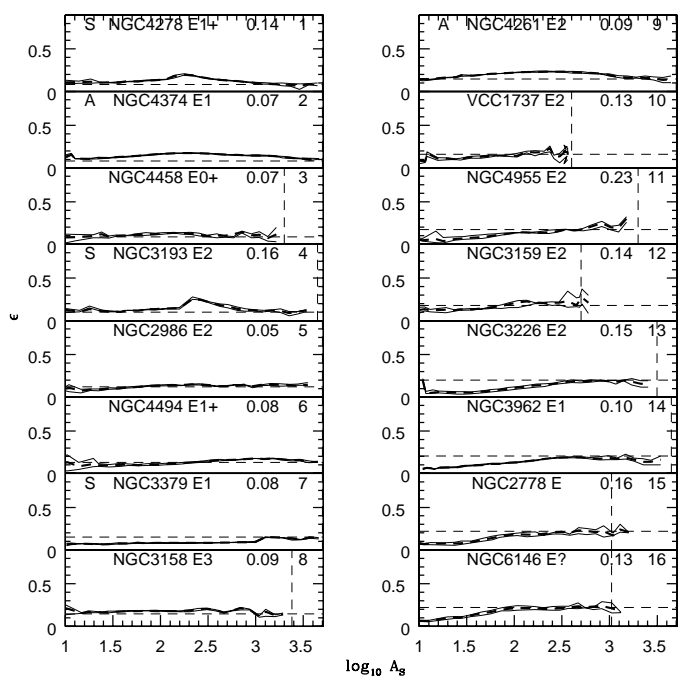}
\caption{Ellipticity as a function of contour area ($A_S$) for elliptical 
galaxies in group 1 (2MASS $\epsilon \leq 0.2$). The thick dashed line 
represents the mean ellipticity of three different NIR bands: $J, H,$ and 
$K_s$. The upper and lower thin solid lines are used to show the maximum 
and minimum ellipticity measured in these three bands. 
The horizontal and vertical dashed lines represent, respectively, the 
2MASS estimate of $K_s$ band $3 \times \sigma_n$ isophote ellipticity 
and the area within the contour corresponding to that  isophote. Note 
that for some galaxies the vertical line goes beyond the horizontal 
range of the figure. The NED galaxy name, its RC3 classification, and 
the overall deviation ($\Delta \epsilon$) in ellipticity are shown at 
the top of each panel. The difference between the highest and lowest 
value of the thick dashed line, in the range $\log_{10} A_S \geq 1.5$, 
is used to measure $\Delta \epsilon$. Similar style is followed for the 
rest of the presentation. For the meaning of the symbols ``A'' and ``S'' 
see text. \label{ellpe1}}
\end{figure}

{\bf Ellipticity:} The difference in the estimate of ellipticity is seen 
only for lenticular and spiral galaxies. The galaxies showing the difference 
comprise $\sim 32\%$ of the entire sample. Here we provide the list of these 
galaxies: 11, 14 in Fig. \ref{ellpe1}; 5, 16 in Fig. \ref{ellpe2}; 
1, 3, 6, 8, 9, 11, and 15 in Fig. \ref{ellps0}; 
1, 6, 8 14, and 16 in Fig. \ref{ellps1}; 
1, 7, 13, 14, and 16 in Fig. \ref{ellps2}; 
3, 4, 8, 11, 13, 14, 15, and 16 in Fig. \ref{ellps3}; 
2, 6, 11, 13, 14, 15, and 16 in Fig. \ref{ellps4}.

\begin{figure}
\epsscale{1.15}
\plotone{\figdir/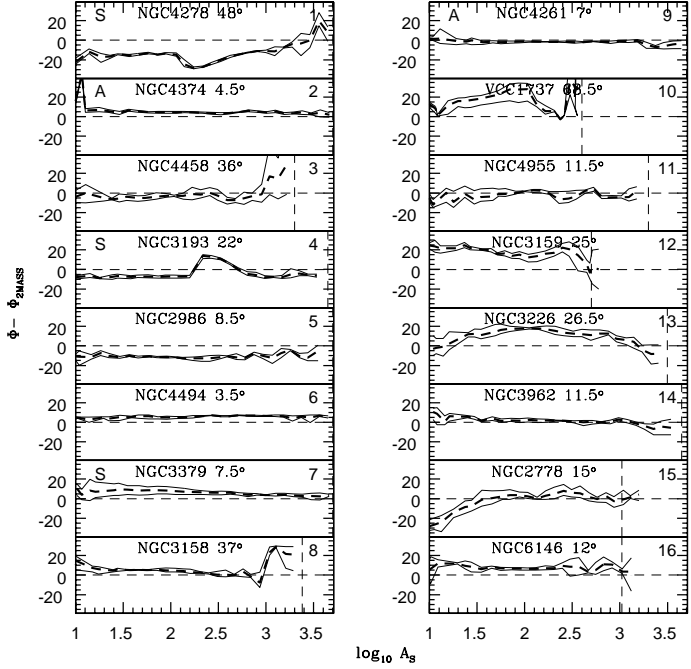}
\caption{Orientations (in degrees) of elliptical galaxies in group 1 
(2MASS $\epsilon \leq 0.2$) presented with respect to the 2MASS $K_s$ 
band $3 \times \sigma_n$ isophote position angle. 
The NED galaxy name and the overall twist ($\Delta \Phi$) in the 
orientation of galaxies are shown at the top of each panel.   
The difference between the highest and lowest value of the thick 
dashed line, in the range, $\log_{10} A_S \geq 1.5$, is used to 
estimate $\Delta \Phi$. \label{ange1}}
\end{figure}

{\bf Orientation:} The difference in the estimate of orientation 
appears in all types of galaxies. For the following galaxies 
($\sim 23\%$ of the entire sample) we have found disagreement: 
3, 8, 10, and 13 in Fig. \ref{ange1}; 
6 , 8, and 9 in Fig. \ref{angs0}; 
2, 6 and 14 in Fig. \ref{angs1}; 
1, 7, 10, 13, 14, and 16 in Fig. \ref{angs2}; 
1, 8, 13, and 15 in Fig. \ref{angs3}; 
1, 4, 8, 9, 11, and 15 in Fig. \ref{angs4}. 
For these galaxies  our estimate of $\Phi$ is either lower or higher than 
the 2MASS. 

We want to point out that while for some galaxies (e. g., 3 and 8 in Fig. 
\ref{ellpe1}) the estimates of $\epsilon$ are in agreement with 2MASS, the 
estimates of $\Phi$ for these galaxies (3 and 8 in Fig. \ref{ange1}) differ. 
In some cases our analyses do not reach the $3 \times \sigma_n$ isophote 
level (e. g., 3 and 8 in Fig. \ref{ellpe1}; 8 and 9 in Fig. \ref{ellps0}; 
6 in Fig. \ref{ellps1}; 7 in Fig. \ref{ellps2}) which is a likely reason 
for disagreement. 

We draw reader's attention to galaxies numbered 11 and 14 in Fig. 
\ref{ellpe1} and listed above. The $\epsilon$ and $\Phi$ profiles of these 
galaxies may give an impression to the reader that they do not reach the 
$3 \times \sigma_n$ level (recall that the figures showing these profiles 
are for smoothed contours). In fact they do reach the level for the 
unsmoothed profiles and disagree with the 2MASS. This particular appearance 
of the profiles arise due to excessive smoothing at the outer part of these 
galaxies (see $\S3$).
 
We provide analyses for smoothed isophotes. Contour smoothing can be thought  
of as a method to minimize the effect of noise to restore the original shape. 
Since the 2MASS results have not been estimated from smoothed isophotes, we 
want to emphasize that there is no guarantee that the shape estimates provided 
by the survey represent the actual shape of galaxy contours. After careful 
comparison of both unsmoothed and smoothed profiles of galaxies, we feel 
reasonably confident that the disagreement is not entirely due to the 
methodological difference. It is likely that our approach is better capable 
of revealing the actual shapes of galaxy isophotes, especially around the 
edge.   
\section{Conclusions}
We have analyzed a sample of galaxies of various Hubble types obtained 
from the 2MASS catalogue. The sample contain 112 galaxies imaged in the 
NIR $J$, $H$, and $K_s$ bands. We have used ellipticity ($\epsilon$) and 
orientation angle ($\Phi$), as functions of area within the isophotal 
contour, as the diagnostic of galaxy shape. These measures have been 
constructed from a set of non-parametric shape descriptors known as the 
Minkowski Functionals (MFs). The ellipticity and orientation for each 
galaxy have been derived at 30 different surface brightness levels in each 
of these bands. The ellipticity and position angle (for $K_s$ band only) 
provided by the 2MASS are used as the reference in our analysis.

\begin{figure}
\epsscale{1.15}
\plotone{\figdir/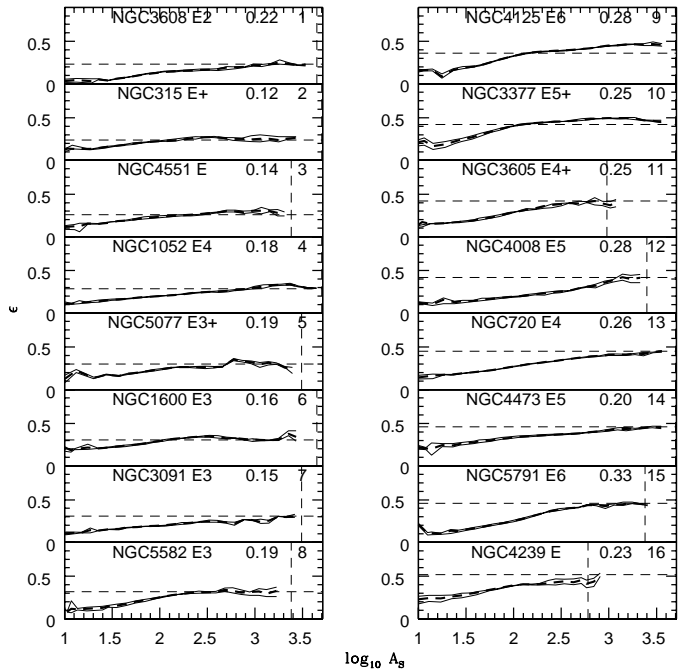}
\caption{Ellipticity as a function of contour area ($A_S$) for 
elliptical galaxies in group 2 (2MASS $\epsilon > 0.2$). 
\label{ellpe2}}
\end{figure}

\begin{figure}
\epsscale{1.15}
\plotone{\figdir/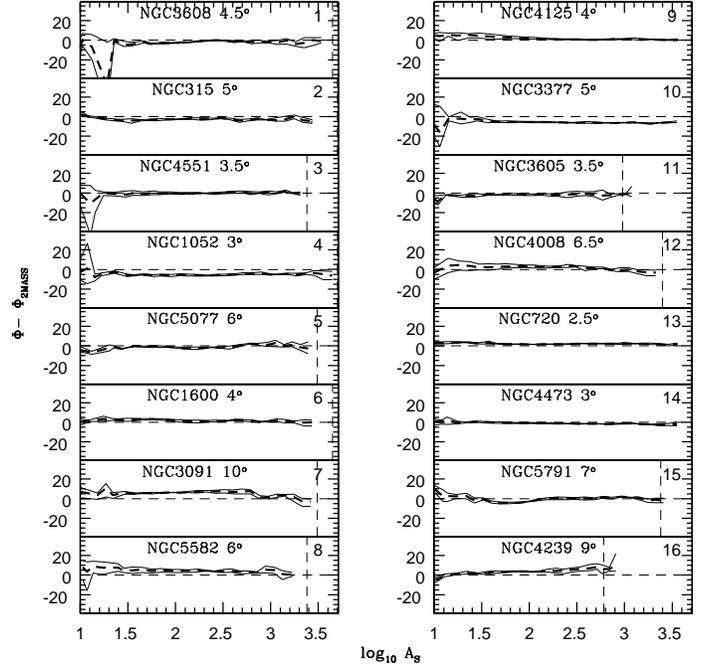}
\caption{The orientations of elliptical galaxies in group 2 
(2MASS $\epsilon > 0.2$). 
\label{ange2}}
\end{figure}

\begin{figure}
\epsscale{1.15}
\plotone{\figdir/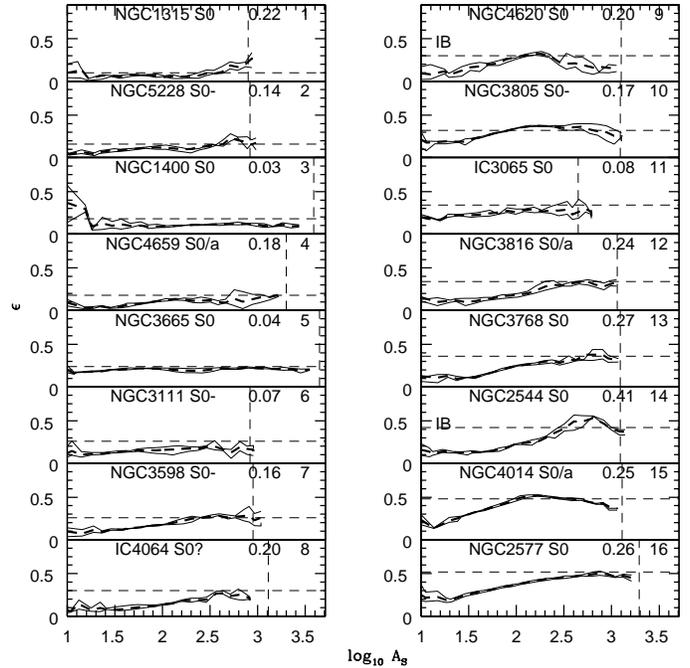}
\caption{Ellipticity as a function of contour area ($A_S$) for 
lenticular galaxies. The label ``IB'' is used to represent 
the galaxy that  has bar like feature only in the infrared. 
\label{ellps0}}
\end{figure}

\begin{figure}
\epsscale{1.15}
\plotone{\figdir/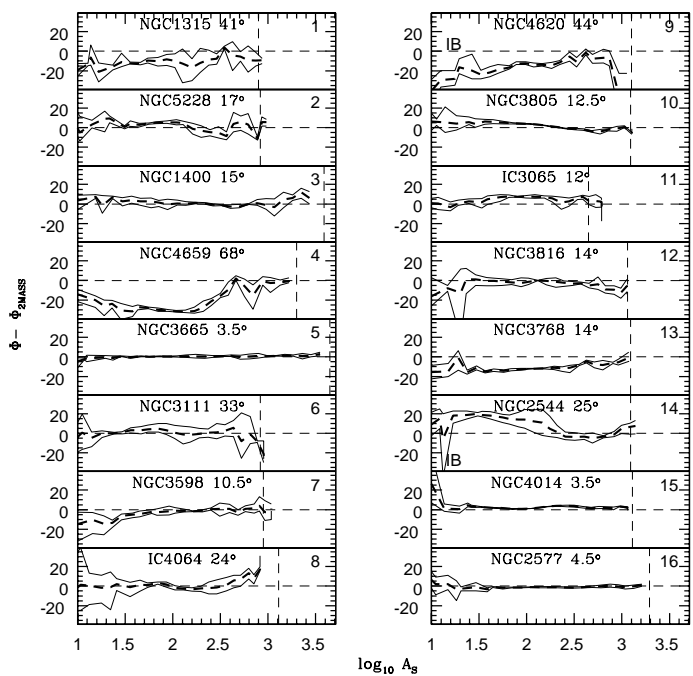}
\caption{Orientations of lenticular galaxies shown in Fig. \ref{ellps0}. 
The label ``IB'' is used to represent the galaxy that has a bar like 
feature only in the infrared. The profiles of galaxy 1 and 4 are scaled 
by a factor 2 to fit the range along the vertical axis. No scaling is 
applied to the other galaxies. \label{angs0}}
\end{figure}

Our results show that the elliptical galaxies with $\epsilon \geq 0.2$ 
appear to be centrally spherical. These galaxies show smooth variations 
in ellipticity with the size, quantified by the area within the contour,  
and are more flattened near the edge than the central region. 
A variation as strong as $\Delta \epsilon \geq 0.25$, from the center 
towards the edge, is noticeable in more flattened systems, e. g., NGC 
4125, NGC 3377, NGC 4008, and NGC 5791 (9, 10, 12, and 15 in group 2 
of elliptical galaxies; Fig. \ref{ellpe2}). This behavior is similar to 
previous studies of ellipticals in visual bands (Jedrzejewski 1987, Fasano 
\& Bonoli 1989, Franx et al. 1989). The similarity in apparent shapes in 
optical and NIR wavelengths indicates that the low ellipticity noticed at 
the central regions of these galaxies is intrinsic rather than an artifact 
contributed by the seeing effect. Additionally the $\epsilon$ profiles of 
these galaxies appear very similar in different NIR bands with a very low 
scatter. This suggests that the morphological differences that are likely 
to appear in different bands are weak in NIR elliptical galaxies.

The twist ($\Delta \Phi$), characterized by the overall change in the 
orientation with radius, observed in the orientations of elliptical 
galaxies decreases with increasing flattening. A relatively small twist 
shown by ellipticals with $\epsilon > 0.2$ suggests that elongation and 
isophotal twist are likely to be anti-correlated in the NIR wave-bands. 
We perform a Spearman correlation test between the twist and the 
deviation in ellipticity ($\Delta \epsilon$) for the entire sample of 
elliptical galaxies. The test result (correlation coefficient -0.35 and 
probability 0.05) indicates a significant anti-correlation between 
$\Delta \Phi$ and $\Delta \epsilon$ (see Fig. \ref{galletta}). Note that 
in the test, a small value of the probability with a negative coefficient 
suggests significant anti-correlation.

To check whether this result is due to any methodological artifact or 
indicates a physical effect, we rerun the test two more times with different 
numbers of galaxies in the elliptical sample. For the first run, we construct 
the sample after removing only galaxy 10 from group 1. 
This galaxy shows the largest twist ($\Delta \Phi \sim 68.5^{o}$) in the 
entire sample and, therefore, we discard it as an outlier. We find that the 
removal of this galaxy slightly weakens the result (correlation coefficient 
-0.34 and probability 0.06) but the correlation still remains  significant. 
In the next run, we remove nine galaxies from the entire sample: the first  
eight galaxies from group 1 because of their spherical shapes 
($\epsilon \sim 0.1$) and galaxy 10. This time the test result gives 
correlation coefficient -0.37 and probability 0.09. Although a greater 
probability indicates less confidence for this subsample, the result does 
not differ substantially from the original sample. It suggests that the 
anti-correlation is a physical effect rather than a fluke. 

The NIR correlation between $\Delta \Phi$ and $\Delta \epsilon$ for 
elliptical galaxies is similar to that in the optical wavelengths. 
For elliptical galaxies in visual bands, it has been known for a while 
that the maximum apparent flattening and the maximum observed twist are 
inversely related (Galletta 1980). An interesting aspect of this 
correlation would be the possible coupling between NIR and optical light 
(Jarrett et al. 2003). 

Elliptical galaxies with $\epsilon < 0.2$ usually show large variation in 
their orientations. These galaxies also show considerable differences both 
in ellipticity and orientation in different bands. The large twist observed 
in these galaxies should be taken with caution since spherical contours 
obtained from nearly spherical galaxies are highly prone to spurious effects 
such as background noise. 

The NIR lenticular galaxies have properties similar to those of ellipticals 
or disk galaxies. The profiles of a few of these galaxies show trends similar 
to ellipticals. A few of these galaxies show characteristic properties 
resembling to the galaxies with bar like structures. 
The lenticular galaxies in our sample, in general, have larger scatter in 
both $\epsilon$ and $\Phi$ than the ellipticals. In the entire sample of 
lenticulars, at least 2 galaxies have bar like signatures in their profiles. 
The observed twists in the lenticulars' orientations are comparable to the 
trend noticed in spherical galaxies ($\epsilon < 0.2$). The properties of 
these galaxies stress the fact that morphological classification strongly 
depends on the wavelength studied. It may be possible that S0 galaxies do 
not exist in the long wavelength part of the spectrum. This is simply our 
speculation and more elaborate studies are needed to make this a definite 
conclusion.

\begin{figure}
\epsscale{1.15}
\plotone{\figdir/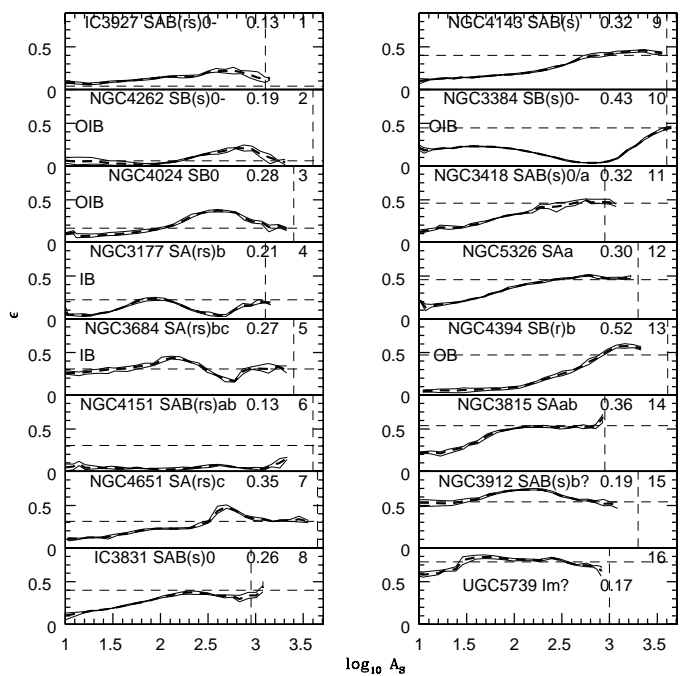}
\caption{Ellipticity as a function of contour area ($A_S$) for spiral 
galaxies in group 1. The labels ``OIB'' and `OB'' represent, 
respectively, ``optical and infrared bar'' and ``optical bar''.
Note that the spiral galaxies are divided into four groups. The groups 
are organized using the scatter ($\delta \epsilon$) observed on the 
ellipticity profiles. The galaxies in spiral group 1 have the least 
scatter ($\delta \epsilon \leq 0.05$). \label{ellps1}}
\end{figure}

The sample of spiral galaxies indicates that a bar like feature is 
ubiquitous in disk galaxies without any significant dependence on 
Hubble's class. This sample has 64 galaxies, of which 24 ($\sim 38\%$) 
are optically-classified barred galaxies. Most of these galaxies show 
features resembling bar(s) in the infrared. We label these galaxies as 
``optical and infrared bar'' (``OIB''). However, we notice something 
interesting. All most one-third of these optical barred galaxies lack 
characteristic features in their NIR profiles. These galaxies include: 
NGC 4394(12) in group 1 (Fig. \ref{ellps1}); NGC 4375(6) and UGC 7073(9) 
in group 3 (Fig. \ref{ellps3}); UGC 2705(4), UGC 3171(9), NGC 5510(13), 
and UGC 1478(16) in group 4 (Fig. \ref{ellps4}).

The sample of spirals include 40 galaxies that are optically classified 
non-barred galaxies. At least 11 of these galaxies show bar like features 
in their NIR profiles, increasing the number of bar like systems to 
$\sim 45\%$.   
Several previous studies have also reported a higher frequency of barred 
systems in disk galaxies (Seiger \& James 1998; Eskridge et al. 2000; 
Laurikainen \& Salo 2002). Our result is in agreement with all of these 
studies except Seiger \& James (1998), who reported $\sim 90\%$ frequency 
of barred galaxies from a sample of 45 spirals imaged in NIR $J$ and $K$ 
bands. Our estimate indicates that a significant fraction of disk galaxies 
are intrinsically non-barred systems. The absence of a bar like feature in 
the profiles of 29 spiral galaxies favors this argument. 
This is also consistent with Eskridge et al. (2000). 

Among 19 normal spiral galaxies, 17 galaxies are SA-type. The other two, 
UGC 5739 in group 1 and IC 2947 in group 3, are irregular/peculiar type 
galaxies. In these 17 SA-type spirals, 5 galaxies have distinct bar like 
features in their shape profiles indicating that $\sim 30\%$ of SA-type 
galaxies have optically hidden bars. Our result is in agreement with 
Eskridge et al. (2000) who reported $\sim 40\%$ galaxies have optically 
hidden bars from a sample of 186 NIR $H$-band spirals containing 51 
non-barred SA-type galaxies. 


Note that several disk galaxies show multiple bar features in their profiles. 
However, in this study we only report their morphologies and do not attempt 
to make any definite conclusion regarding their projected structure since 
identifying galaxies with multiple bars is a complicated issue (see Wozniak 
et al. 1995; Erwin \& Sparke 1999). It is important to note that our estimate 
of the frequency of barred galaxies is done by visual inspection. Since the 
sample of galaxies used in this study is not based on rigorous selection 
criteria, the results should be taken as an approximate estimate.

\begin{figure}
\epsscale{1.15}
\plotone{\figdir/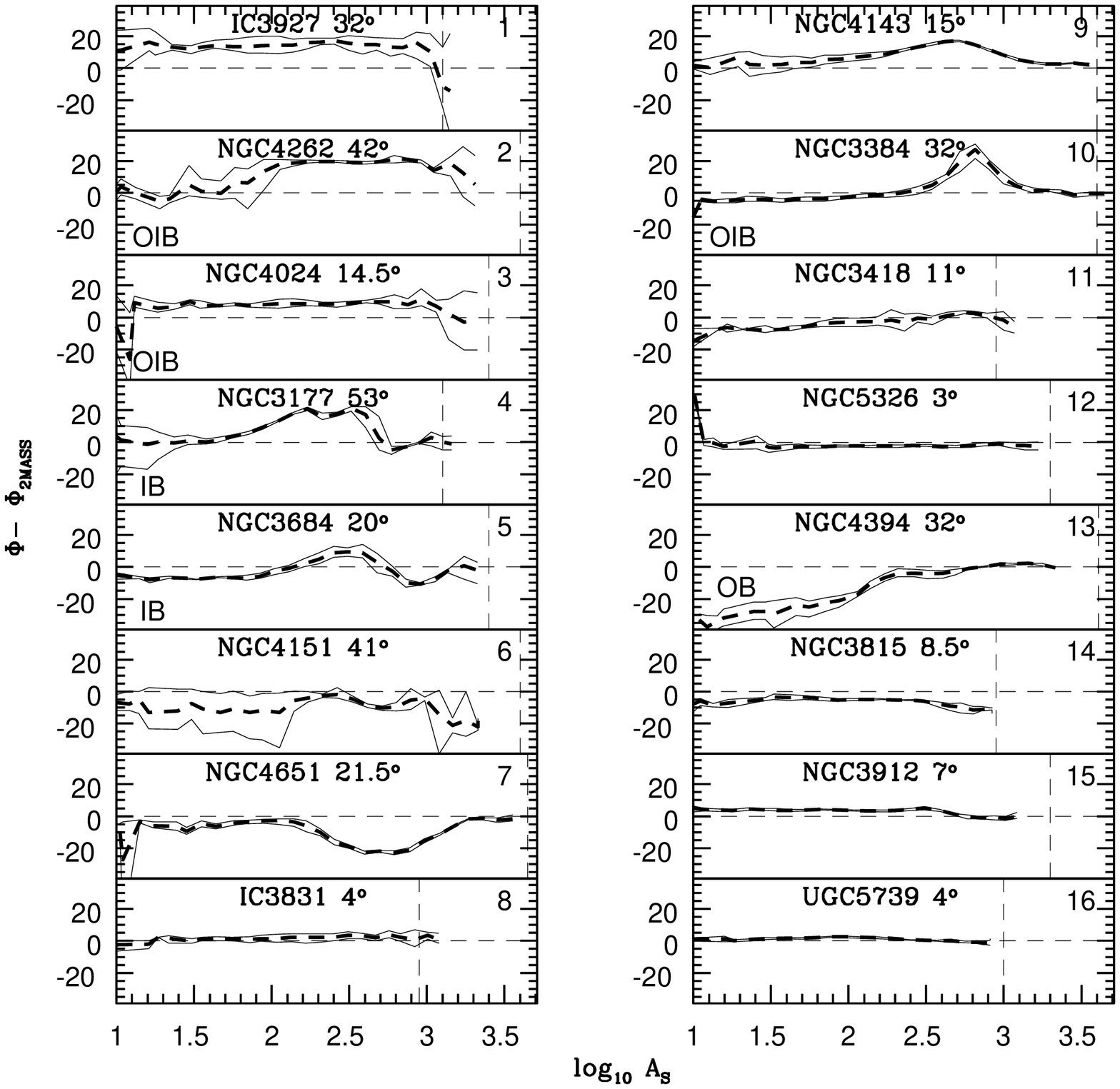}
\caption{Orientations of galaxies in spiral group 1. Galaxies,  
numbered as 2, 4, and 6, are scaled by a factor 2 to fit the 
range along the vertical axis. No scaling is applied to other 
galaxies. 
\label{angs1}}
\end{figure}

\begin{figure}
\epsscale{1.15}
\plotone{\figdir/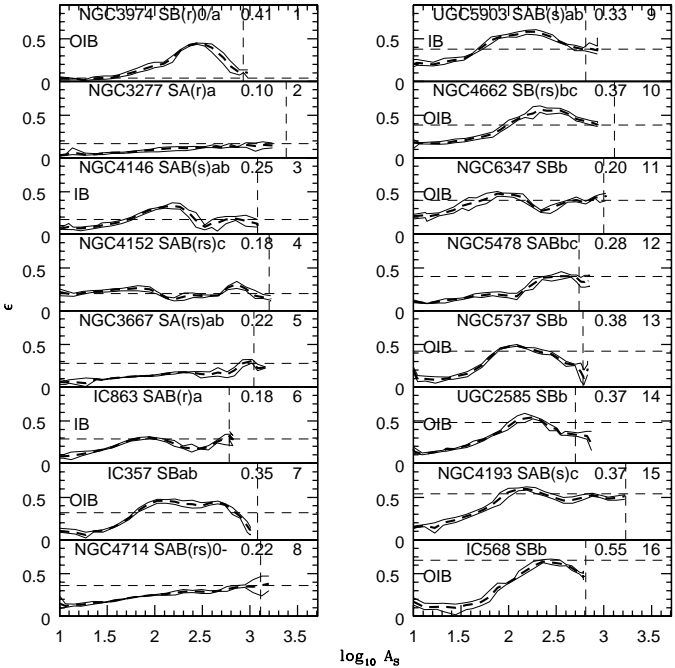}
\caption{Ellipticity as a function of contour area ($A_S$) for spiral 
galaxies in group 2. The galaxies have scatter in ellipticity in the 
range $0.05 < \delta \epsilon \leq 0.1$ when $J,$ $H,$ and $K_s$ band 
measurements are compared.
\label{ellps2}}
\end{figure}

\begin{figure}
\epsscale{1.15}
\plotone{\figdir/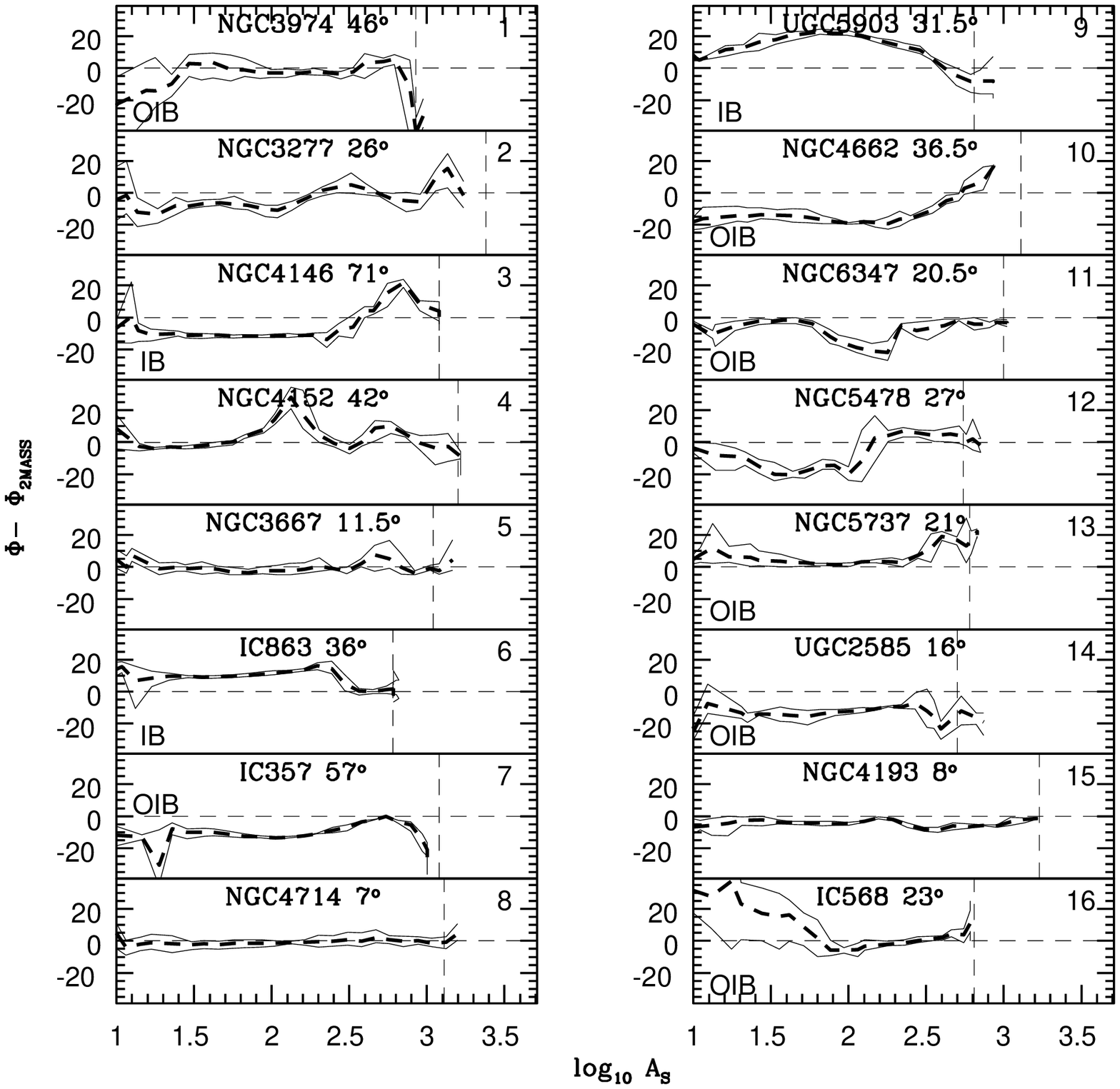}
\caption{Orientations of galaxies in spiral group 2. Galaxies 
numbered as 3, 6, and 7, are scaled by a factor 2 to fit the range 
along the vertical axis. No scaling is applied to other galaxies. 
\label{angs2}}
\end{figure}

\begin{figure}
\epsscale{1.15}
\plotone{\figdir/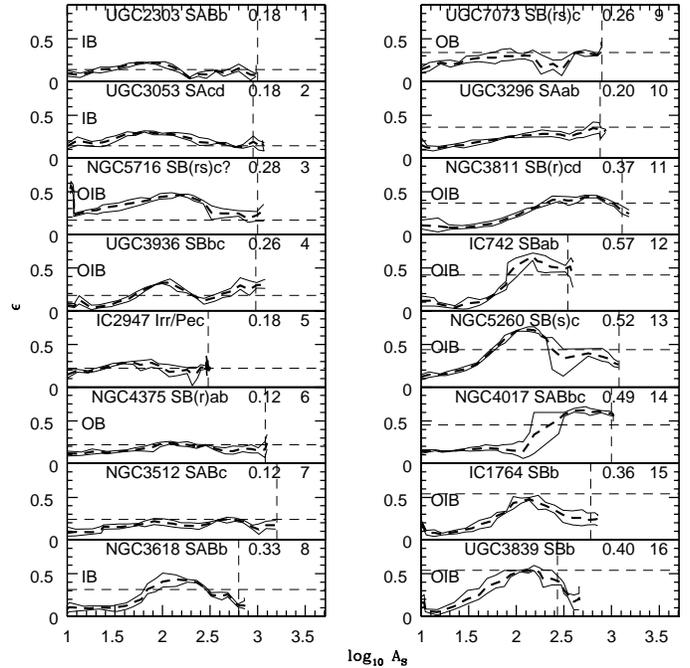}
\caption{Ellipticity as a function of contour area ($A_S$) for spiral 
galaxies in group 3. The galaxies have scatter in the range, 
$0.1 < \delta \epsilon \leq 0.2$, in $J, H,$ and $K_s$ band 
measurements. Note that NGC 4375(6), an optical barred system (``OB'') 
lacks characteristic feature in its NIR profile. 
\label{ellps3}}
\end{figure}

\begin{figure}
\epsscale{1.15}
\plotone{\figdir/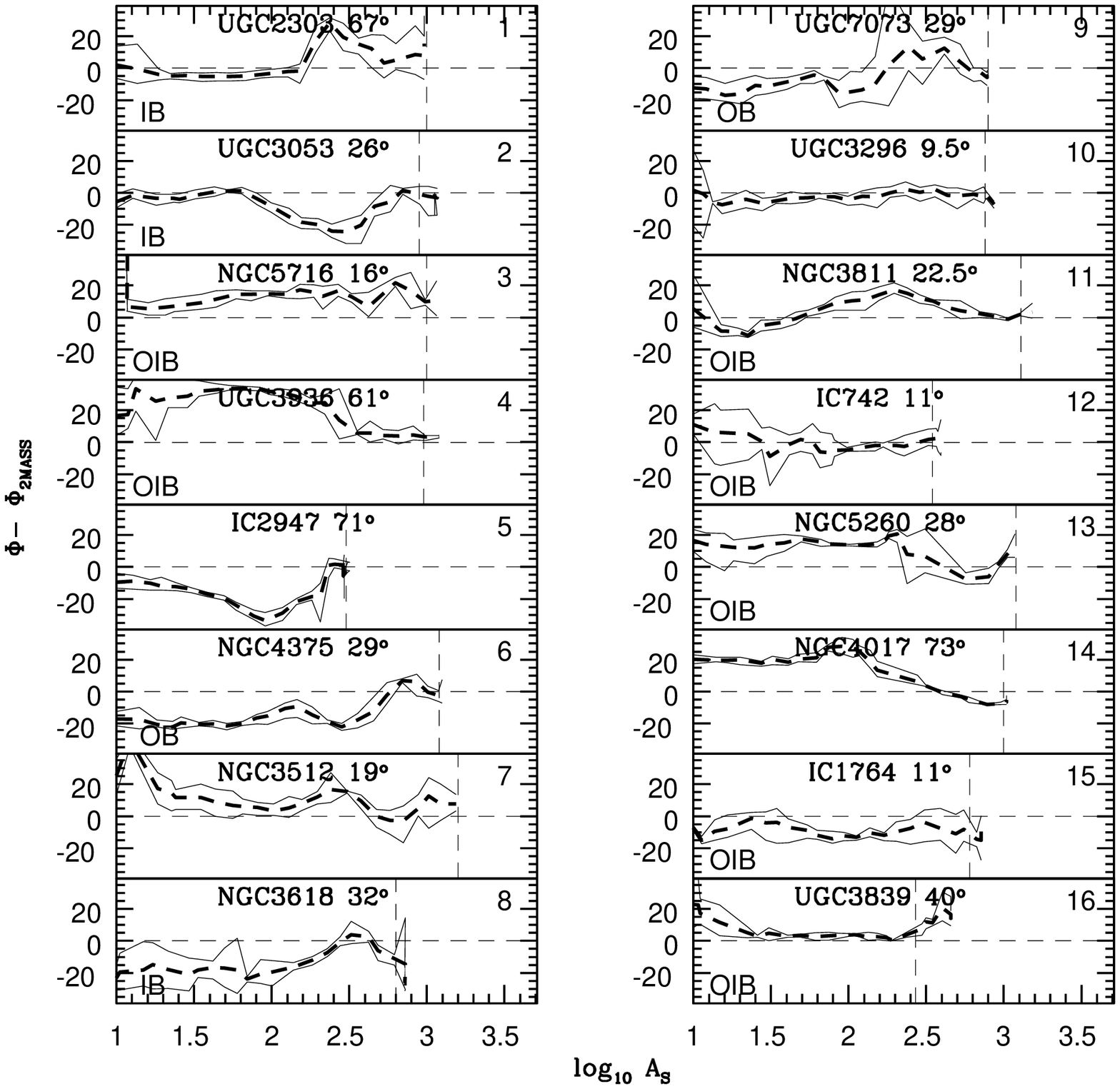}
\caption{Orientations of galaxies in spiral group 3. Galaxies,  
numbered as 1, 4, 5, 14, and 16, are scaled by a factor 2 to 
fit the range along the vertical axis. No scaling is applied to 
other galaxies. \label{angs3}}
\end{figure}

\begin{figure}
\epsscale{1.15}
\plotone{\figdir/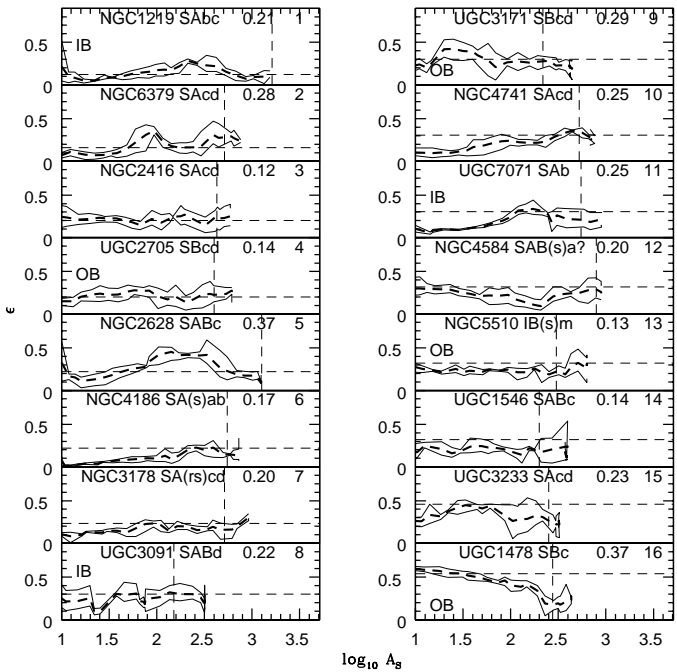}
\caption{Ellipticity as a function of contour area ($A_S$) for spiral 
galaxies in group 4. The galaxies have scatter $\delta \epsilon > 0.2$ 
in $J, H,$ and $K_s$ band measurements. 
\label{ellps4}}
\end{figure}

\begin{figure}
\epsscale{1.15}
\plotone{\figdir/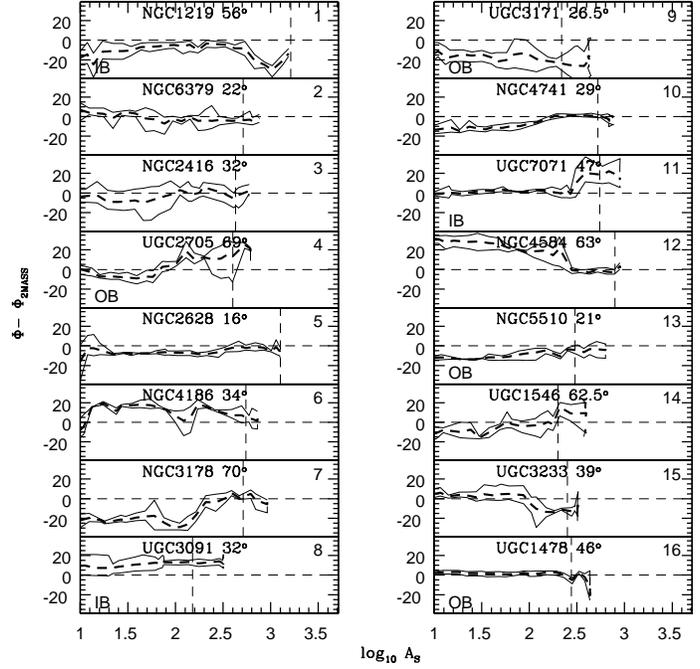}
\caption{Orientations of galaxies in spiral group 4. Galaxy UGC 3091(8) 
is scaled by a factor 4 to fit the range along the vertical axis. The other 
galaxies are scaled by a factor 2.\label{angs4}}
\end{figure}

\begin{figure}
\epsscale{0.99}
\plotone{\figdir/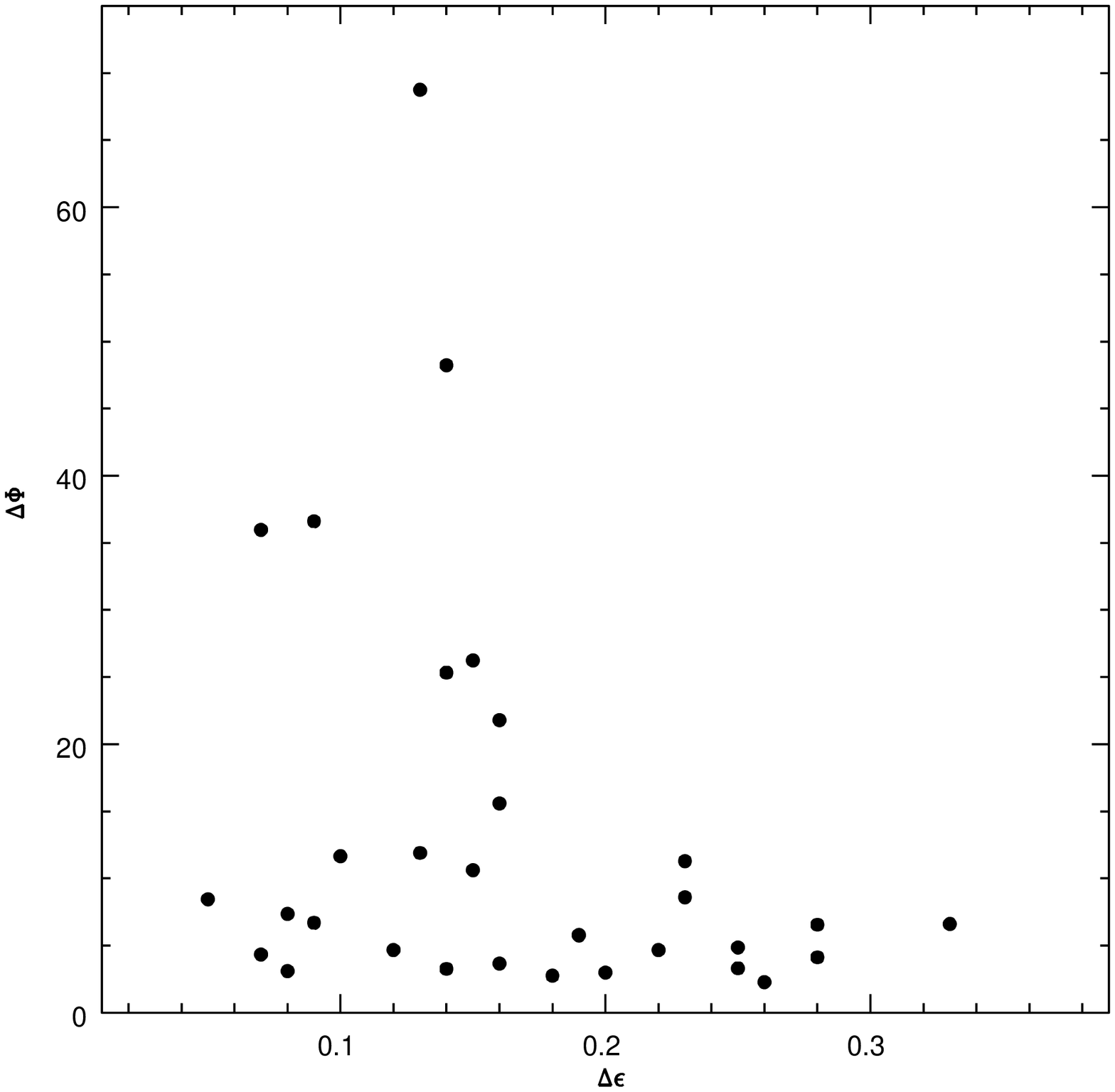}
\caption{Twist ($\Delta \Phi$, in degree) versus deviation in ellipticity 
($\Delta \epsilon$) for the entire sample of elliptical galaxies. The Spearman 
correlation test, with a correlation coefficient of $-0.35$ and a probability 
of $0.05$, indicates a significant anti-correlation between these two 
parameters. \label{galletta}}
\end{figure}

The overall isophotal twists observed in the orientations of spiral galaxies 
range between $\sim 3^{o}-71^{o}$. Many NIR barred spirals appear to have 
smaller but notable twists in their orientations than the twist at optical 
wavelengths (see e. g., Wozniak et al. 1995). Two conclusions can be drawn 
after analyzing the orientation profiles of the NIR images. First, the NIR 
light in $J, H,$ and $K_s$ bands is not fully decoupled from Population I 
light. They are most likely weakly coupled. Second, the different regions 
(central bulge, bar, and outer disk) of disk galaxies are dynamically linked. 
It is manifested in the continuous change in the orientations of the majority 
of the disk galaxies. Exceptions are noticed in few cases, e. g., NGC 3384 
(10, group 1; Fig. \ref{angs1}), NGC 4152 (4, group 2; Fig. \ref{angs2}), 
UGC 2303 (1, group 3; Fig. \ref{angs3}), where the change in the orientation 
is very sharp. But it is important to note that in the region where the sharp 
changes are observed, the galaxy contours appear to be very spherical 
($\epsilon \leq 0.15$). A drastic change in the orientations of spherical 
contours does not indicate reliably the actual nature of the internal 
structure since these types of contours are prone to spurious effects.   
\\
\\
\noindent{\bf Acknowledgments}
We thank the referee for his constructive criticisms and many 
helpful suggestions. We are indebted to Bruce Twarog for thorough 
and critical reading of the manuscript. His suggestions improved the 
paper significantly. 
NR thanks Tom Jarrett, Hume Feldman, and Barbara Anthony-Twarog for 
useful discussions. NR acknowledges the GRF support from the University 
of Kansas in 2002-2003. SFS acknowledges the support from the Nonlinear 
Cosmology Program, 2003 at Observatory de la Cote d'Azur in Nice, France 
during the summer 2003 and the AAS travel grant.  

This research has made use of the NASA/IPAC Infrared Science Archive, 
which is operated by the Jet Propulsion Laboratory, California Institute 
of Technology, under contract with the National Aeronautics and Space 
Administration.

\section{Appendix}
Here we briefly describe the role of contour smoothing in our analyses. We 
also discuss the sensitivity of the parameters to any disturbance present 
on the contour. 

\begin{figure}
\epsscale{1.1}
\plotone{\figdir/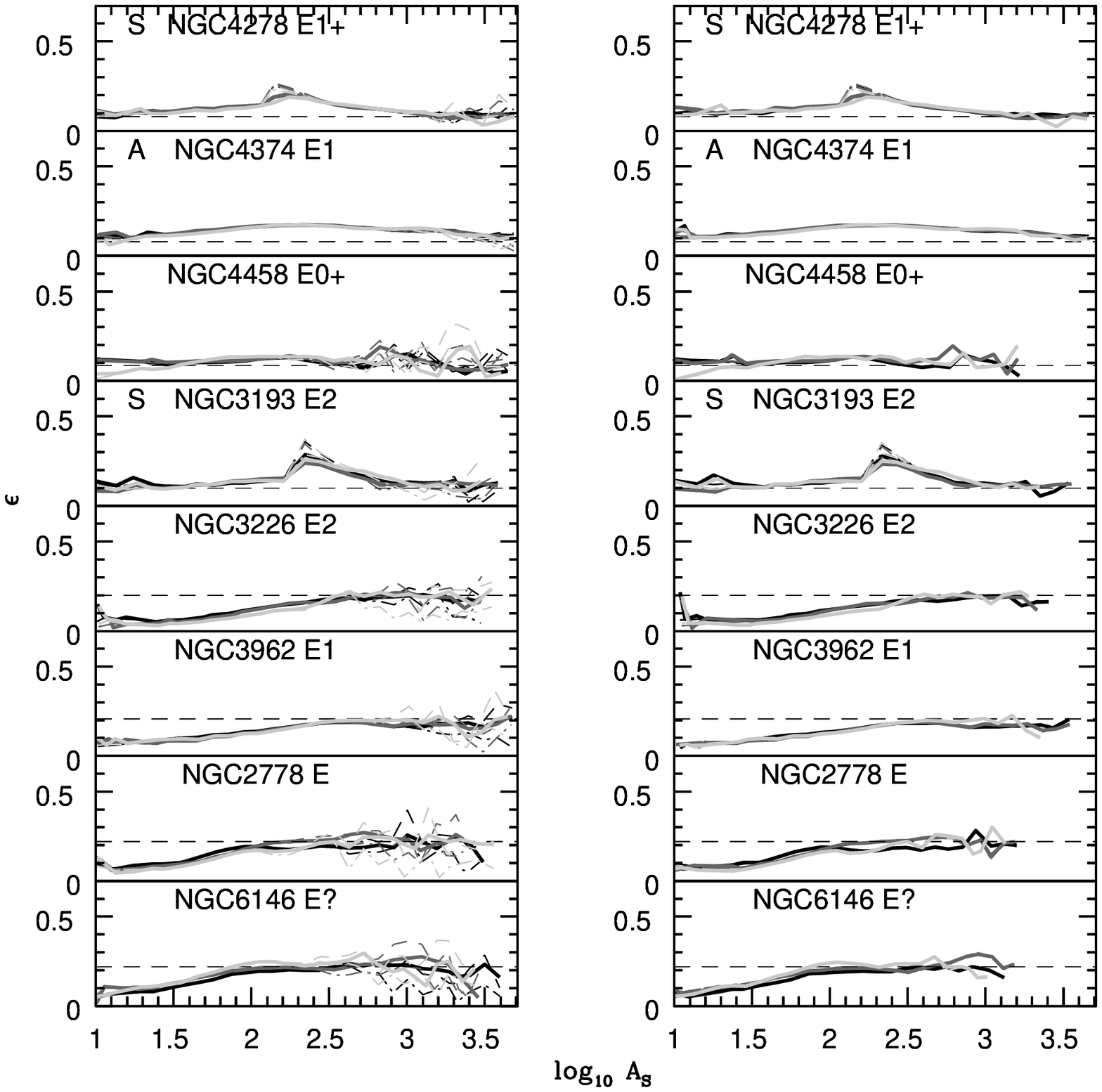}
\caption{Ellipticity  as a function of area ($A_S$) obtained from 
unsmoothed (left panels) and smoothed (right panels) contours of 
a selection of elliptical galaxies from group 1. 
The dark, medium, and light colors show the measurements from $J,$ 
$H,$ and $K_s$ bands respectively. For each band the solid, dashed 
and dashed-dotted lines represent the ellipticity measure of the 
area, perimeter and EC AEs. The symbols ``A'' and ``S'' are explained 
in the text.
Notice how contour smoothing reduces the effect of noise, retaining 
the main features of the contours; different AEs for 
each band converge and stay on top of each other. The convergence 
indicates that the galaxy contours are indeed elliptical in nature, 
revealed by smoothing. 
\label{ellp_all_e1_unsm}}
\end{figure}

\begin{figure}
\epsscale{1.1}
\plotone{\figdir/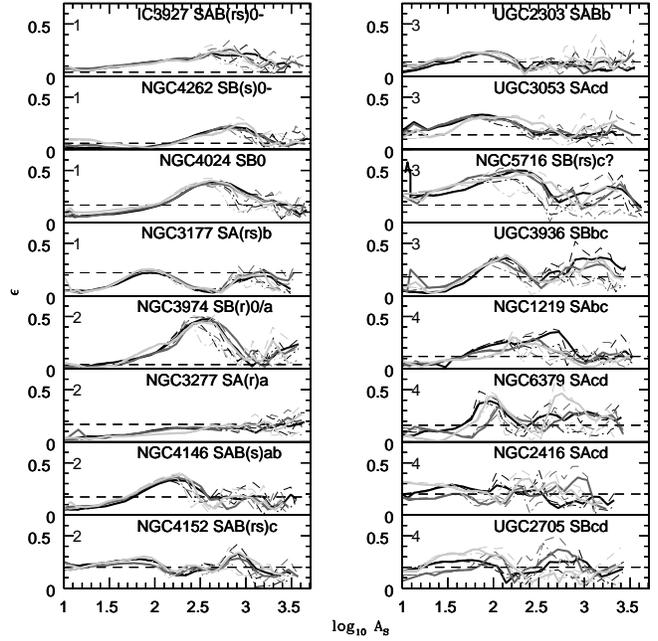}
\caption{Ellipticity  as a function of area ($A_S$) obtained from 
unsmoothed contours of spiral galaxies. 
The first four galaxies from each spiral group are shown in this 
figure marked by numbers 1 to 4, where 1 stands for group 1 and so on. 
The line styles and colors are similar to Fig. \ref{ellp_all_e1_unsm}. 
\label{ellp_all_spi_unsm}}
\end{figure}

\begin{figure}
\epsscale{1.1}
\plotone{\figdir/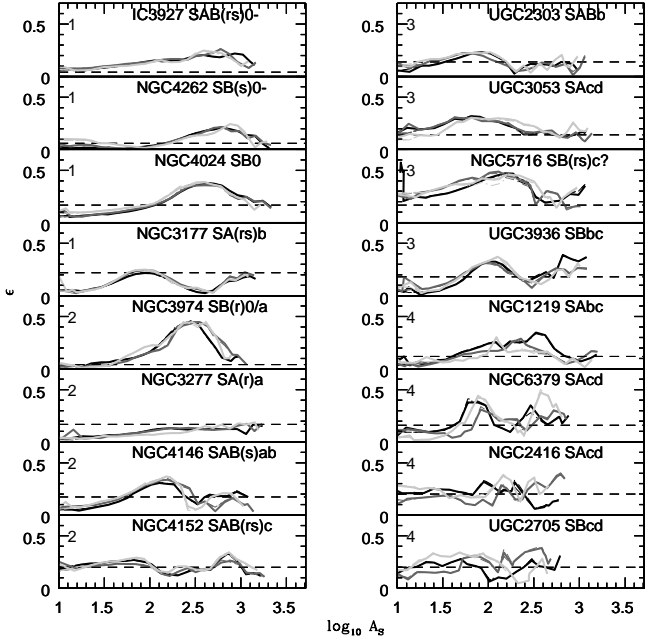}
\caption{Ellipticity  as a function of area ($A_S$) for smoothed 
contours of spiral galaxies in Fig. \ref{ellp_all_spi_unsm}. 
The line styles and colors are the same as before. 
The convergence of the AEs is a result of smoothing 
and illustrates the fact that the contours of the spiral galaxies are 
elliptic in nature although the ellipticity varies with image size. 
In comparison to Fig. \ref{ellp_all_spi_unsm}, one can see that 
contour smoothing reduces noise, keeping the main features intact.   
One can also notice that the color difference is stronger in late 
type spirals (lower four panels on the right, marked with the number 
4). \label{ellp_all_spi_sm}}
\end{figure}

In order to emphasize the role of contour smoothing we show in Fig. 
\ref{ellp_all_e1_unsm} the ellipticity measured from all three AEs, both 
for unsmoothed and smoothed elliptical galaxies in group 1. In this figure 
the dark, medium, and light colors are used to represent the $J, H,$ and 
$K_s$ band respectively. The solid, dashed, and dashed-dotted lines show 
the area, perimeter, and EC ellipses for each band. As we can see from the 
unsmoothed contours in the left panel, the ellipses differ from one another 
in each band mostly near the edge. This deviation is caused by the background 
noise since the galaxy surface brightness distribution is steeper near the 
center and shallower outward. We should also note that in a region far 
away from the center of a galaxy the absolute value of the distribution is 
very small. In either case any kind of background noise will have a strong 
effect on the outer part of the distribution. 
Therefore when one constructs contours along the edge of a galaxy, they 
appear highly deviated no matter what their true shapes are. If the 
contours were truly elliptical, they could appear in any arbitrary shape 
depending on the amount distortion. Since the contours are not elliptic 
anymore, the AEs diverge from one another. This divergence reflects the 
apparent shape of the contours. One can see from the right panel that contour 
smoothing reduces the role of noise significantly. It helps restore the 
original shape of the contours. At the same time one can also notice that the 
differences among the ellipses reduce substantially. To demonstrate further, 
we collect first four galaxies from each group of spiral galaxies and show the 
ellipticity for both unsmoothed and smoothed contours in Figs. 
\ref{ellp_all_spi_unsm} and \ref{ellp_all_spi_sm}, respectively. The motive, 
once again, is to reemphasize the role of contour smoothing. 

When we compare Figs. \ref{ellp_all_e1_unsm} and \ref{ellp_all_spi_sm}, it 
appears that the color difference is, in general, stronger for spiral 
galaxies and within spirals, it is stronger for certain Hubble types. We 
see that the galaxies in spiral group 4, which are mostly late-type, have 
significant differences in their shapes (at least what is revealed by the 
$\epsilon$ parameter) when studied in three different NIR bands.   

\begin{figure}
\epsscale{0.99}
\plotone{\figdir/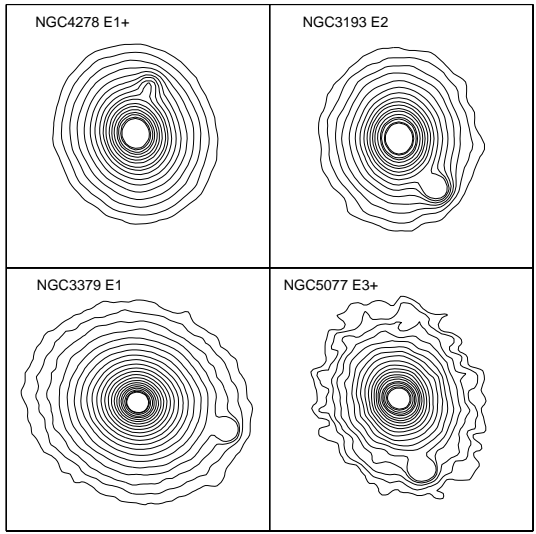}
\caption{The unsmoothed contour plot of galaxies in $J$ band. In each of 
these galaxy images a foreground star is left embedded. The presence of 
the lobe-like feature on contours is a signature of embedded foreground 
star.\label{star_cont}}
\end{figure}

We now proceed to check the sensitivity of the Minkowski shape measures to 
detect image contamination by a foreground star. The $J$ band contours of 
NGC 4278, NGC 3193, NGC 3379 (1, 4, and 7 in elliptical group 1), are shown 
in Fig. \ref{star_cont}. This figure also includes the contours of NGC 5507 
(5 in elliptical group 2) from the same band. Note that the presence of a 
star is apparent in all three bands. It is most prominent in the $J$ band 
and this why we show the contour plot in this band.

\begin{figure}
\epsscale{1.1}
\plotone{\figdir/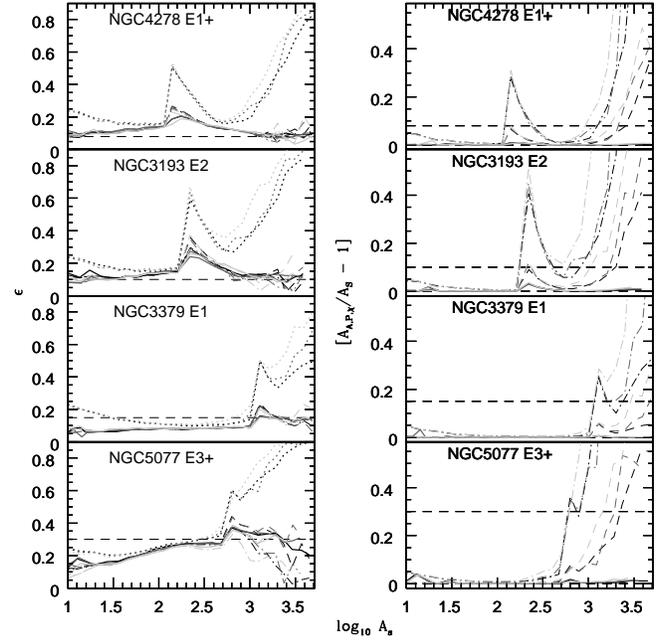}
\caption{Ellipticity (left panel) and relative differences in the areas 
enclosed by different AEs (right panel) for elliptical galaxies where a 
foreground star is embedded in the galaxy images. The figure shows 
information from all three bands. Dark, medium, and gray colors represent, 
respectively, the $J, H,$ and $K_s$ band.
The dotted line represents ellipticity from the scalar functional (see 
appendix for more); the solid, dashed, and dashed-dotted line represent 
the ellipticities of the area, perimeter, and EC AEs, respectively. 
A sharp kink in the ellipticity profiles is the signature of the embedded 
foreground star. A similar feature can also be seen from the plot showing 
the ratios of the sizes of AEs. 
\label{star_measure}}
\end{figure}

The left panel of Fig. \ref{star_measure} shows the ellipticity profiles of 
these galaxies and the right panel shows the relative difference in areas 
enclosed by three tensor AEs. In the left panel we use solid, dashed and 
dashed-dotted lines  represent the measures from area, perimeter and EC AEs. 
The dotted line is the measure from the scalar functional AE that has not 
been used in this analysis. It is included only for the purpose of 
illustration. Note that the construction of the scalar ellipse is different 
from the tensor ellipses. Since the scalar functionals are simply the area 
($A_S$) and perimeter ($P_S$) of a given contour, the scalar AE is an ellipse 
that has the same area and perimeter as the contour itself. 
For a perfect elliptical contour, the scalar AE will converge with its tensor 
counterparts. The scalar functionals do not have enough information to 
attribute an orientation to its ellipse. The right panel does not show any 
measurement from the scalar functional. 

From Figs. \ref{star_cont} and \ref{star_measure}, we see that a foreground 
star embedded in an image causes galaxy isophotes to deviate  from their 
original shapes. It usually adds small, lobe-like features to the otherwise 
smoothed and spherical contours. The functionals easily pick up this type of 
signal present on the contour and translate it to the shape parameters. 
This demonstration highlights the fact that MFs based measures can be used 
for automatic detection of features attached to the image body. 

Our analysis is further supported by NGC 5077 (number 5 in elliptical group 
2). This galaxy is interesting for our purpose since it has been archived 
in the 2MASS catalogue as an image free from foreground star. 
The contour plot of the galaxy, however, reveals that it is not the case 
since the isophotal contours are perturbed by an uncommon feature. Apart from 
the contour plot, the behavior of the ellipticity profile or the plot showing 
ratios of the sizes of AEs also give strong indication of the presence of 
something unusual in the image body. 
From the extent of the feature along the radial direction one can infer the 
possible nature of the object that distorts the image contours. A close 
inspection of the contours and a comparison with the other galaxies mentioned 
above lead us to conclude that a foreground star is still left embedded in 
the galaxy image. 

With these examples at hand, we feel confident to suggest that both 
ellipticity and the ratios of the sizes of AEs can be used simultaneously as 
filtering tools in image processing. These measures may appear useful to 
reduce the chance of contamination by foreground star while constructing large 
galaxy catalogues from surveys such as Sloan Digital Sky Survey (SDSS).

\newpage

\begin{thebibliography}{}
\bibitem[\protect\citename{Abraham \etal} 1999]
{abraham-etal99}
Abraham R. G., Merrifield M. R., Ellis R. S., 
Tanvir N. R., Brinchmann J., 1999, MNRAS, 308, 569 

\bibitem[\protect\citename{Abraham \&  Merrifield} 2000]
{abr-mer99}
Abraham R. G., Merrifield M. R., 2000, ApJ, 120, 2835

\bibitem[\protect\citename{Athanassoula \etal} 1990]
{athanassoula-etal90}
Athanassoula E., Morin S., Wozniak H., Puy D., 
Pierce M. J., Lombard J., Bosma A., 1990, MNRAS, 245, 130

\bibitem[\protect\citename{Bender \& M\"{o}llenhoff} 1987]
{ben-mol87}
Bender R.,  M\"{o}llenhoff C., 1987, A\&A, 177, 71

\bibitem[\protect\citename{Beisbart} 2000]{beisbart00}
Beisbart C., 2000, Ph.D. Thesis, 
Ludwig-Maximilians-Universit\"{a}t, M\"{u}nichen, Germany

\bibitem[\protect\citename{Beisbart, Buchert, \& Wagner} 2001]
{beisbart-etal01a}
Beisbart C., Buchert T., Wagner H., 2001a, Physica A, 293, 592B

\bibitem[\protect\citename{Beisbart, Valdarnini \& Buchert} 2001]
{beisbart-etal01b}
Beisbart C., Valdarnini R., Buchert T., 2001b, A\&A, 379, 412


\bibitem[\protect\citename{Buta\& Block} 2001]{but-blo01}
Buta R., Block D. L., 2001, ApJ, 550, 243  


\bibitem[\protect\citename{Carter} 1978]{carter78}
Carter D., 1978, MNRAS, 182, 797 

\bibitem[\protect\citename{Carter \& Metcalfe} 1980]{car-met80}
Carter D., Metcalfe N., 1980, MNRAS, 191, 325

\bibitem[\protect\citename{de Vaucouleurs \etal} 1991]{dev-etal91}
de Vaucouleurs G., de Vaucouleurs A., Corwin H. G. Jr., Buta R., 
Paturel G., Fouqu\'{e} P., 1991, Third Reference Catalogue 
of Bright Galaxies (Berlin: Spinger) (RC3)  

\bibitem[\protect\citename{Eskridge \etal} 2000]{esk-etal00}
Eskridge P. B. et al., 2000, AJ, 119, 536 

\bibitem[\protect\citename{Erwin \& Sparke} 1999]{erw-spa89}
Erwin P, Sparke L. S., 1999, ApJ, 521, L37

\bibitem[\protect\citename{Erwin \& Sparke} 2003]{erw-spa03}
Erwin P, Sparke L. S., 2003, ApJSS, 146, 299

\bibitem[\protect\citename{Fasano \& Bonoli} 1989]{fas-bon89}
Fasano G., Bonoli C., 1989, A\&ASS, 79, 291

\bibitem[\protect\citename{Franx, Illingworth \& Heckman} 1989]
{fra-etal89}
Franx M., Illingworth G., Heckman T., 1989, AJ, 98, 538

\bibitem[\protect\citename{Galletta} 1980]{galletta80}
Galletta G., 1980, A\&A, 81, 179

\bibitem[\protect\citename{Hobson, Jones, \& Lasenby} 1999]
{hobson-etal99}
Hobson M. P., Jones A. W., Lasenby A. N., 1999, MNRAS, 309, 125

\bibitem[\protect\citename{Jarrett} 2000]{jarrett00}
Jarrett T. H., 2000, PASP, 112, 1008

\bibitem[\protect\citename{Jarrett \etal} 2000]{jarrett-etal00}
Jarrett T. H., Chester T., Cutri R., Schneider S., Skrutskie R.,
Huchra J. P., 2000, ApJ, 119, 2498

\bibitem[\protect\citename{Jarrett \etal} 2003]{jarrett-etal03}
Jarrett T. H., Chester T., Cutri R., Schneider S., Skrutskie R.,
Huchra J. P., 2003, AJ, 125, 525

\bibitem[\protect\citename{Jedrzejewski } 1987]{jedrzejewski87}
Jedrzejewski R. I., 1987, in Structure and Dynamics of Elliptical 
Galaxies, ed. T. de Zeeuw, (Dordrecht: Reidel), p. 37 

\bibitem[\protect\citename{Kerscher \etal} 1997]{kerscher-etal97}
Kerscher M. et al., 1997, MNRAS, 284, 73

\bibitem[\protect\citename{Kerscher \etal} 2001a]
{kerscher-etal01a}
Kerscher M., Mecke K., Schmalzing J., Beisbart C., Buchert T.,
Wagner H., 2001a, A\&A, 373, 1

\bibitem[\protect\citename{Kerscher \etal} 2001b]
{kerscher-etal01b}
Kerscher M. et al., 2001b, A\&A, 377, 1

\bibitem[Leach 1981]{leach81}
Leach R., 1981, ApJ, 248, 485

\bibitem[Lauer 1981]{lauer85}
Lauer T., 1985, MNRAS, 216, 429 

\bibitem[\protect\citename{Laurikainen, Salo, \& Rautiainen} 2002]
{laurikainen-etal02}
Laurikainen E., Salo H., Rautiainen P., 2002, MNRAS, 331, 880

\bibitem[\protect\citename{Laurikainen \& Salo} 2002]
{lau-sal02}
Laurikainen E., Salo H., 2002, MNRAS, 337, 1118

\bibitem[\protect\citename{Michel-Dansac \& Wozniak} 2004]
{mic-woz04}
Michel-Dansac L., Wozniak H., 2004, A\&A, 421, 863

\bibitem[Martin 1995]{martin95}
Martin P., 1995, AJ, 109, 2428

\bibitem[\protect\citename{Mecke, Buchert, \& Wagner} 1994]
{mecke-etal94}
Mecke K. R., Buchert T., Wagner H., 1994, A\&A, 288, 697

\bibitem[\protect\citename{Minkowski}  1903]{min03}
Minkowski H., 1903, Math. Ann., 57, 447


\bibitem[\protect\citename{Novikov, Feldman, \& Shandarin} 1999]
{novikov-etal99}
Novikov D., Feldman H. A., Shandarin S. F., 1999,
Int. J. Mod. Phys., D8, 291

\bibitem[\protect\citename{Novikov, Schmalzing, \& Mukhanov} 2000]
{novikov-etal00}
Novikov D., Schmalzing J., Mukhanov V. F., 2000, A\&A, 364, 17

\bibitem[\protect\citename{Peletier \etal} 1990]{pel-etal90}
Peletier, R. F., Davis, R. L., Illingworth, G. D., Davis, L. E., 
Cawson M., 1990, AJ, 100, 1091  

\bibitem[\protect\citename{Quillen, Frogel \& Gonzalez} 1994]
{qui-etal94} 
Quillen A. C., Frogel J. A., Gonzalez R. A, ApJ, 1994, 437, 162

\bibitem[\protect\citename{Rahman \& Shandarin} 2003]
{rah-sh03}
Rahman N., Shandarin S. F., 2003, MNRAS, 343, 933 (paper I)

\bibitem[\protect\citename{Rozas, Knapen \& Beckman} 1998]
{roz-etal98}
Rozas M., Knapen J. H., Beckman J. E., 1998, MNRAS, 301, 631
 
\bibitem[\protect\citename{Sahni \etal} 1998]{sah-etal98}
Sahni V., Sathyaprakash B. S., Shandarin S. F., 1998, 
ApJ, 495, L5
 
\bibitem[Schmalzing \& Buchert 1997]{sch-buc97}
Schmalzing J., Buchert T., 1997, ApJ, 482, L1

\bibitem[Schmalzing \& Gorski 1998]{sch-gor98}
Schmalzing J., Gorski K. M., 1998, MNRAS, 297, 355

\bibitem[\protect\citename{Schmalzing \etal} 1999]
{schmalzing-etal99}
Schmalzing J., Buchert T., Melott A. L., Sahni V.,
Sathyaprakash B. S., Shandarin S. F., 1999, ApJ, 526, 568

\bibitem[\protect\citename{Seigar \& James} 1998]{sei-jam98a}
Seigar M. S., James P. A., 1998, MNRAS, 299, 672

\bibitem[\protect\citename{Shandarin} 2002]{shandarin02}
Shandarin S. F., 2002, MNRAS, 331, 865

\bibitem[\protect\citename{Shandarin \etal} 2002]
{shandarin-etal02}
Shandarin S. F., Feldman H. A., Xu Y., Tegmark M., 
2002, ApJS, 141, 1

\bibitem[\protect\citename{Sheth \etal} 2003]{she-etal03}
Sheth J., Sahni V., Shandarin S. F., Sathyaprakash B. S., 
2003, MNRAS, 343, 22S

\bibitem[\protect\citename{Williams \& Schwarzschild} 1979]
{wil-sch79}
Williams T. B., Schwarzschild M., 1979, ApJ, 227, 56


\bibitem[\protect\citename{Wozniak \& Pierce} 1991]{woz-pie91}
Wozniak H., Pierce M. J., 1991, A\&AS, 88, 325

 
\bibitem[\protect\citename{Wozniak \etal} 1995]{wozniak-etal95}
Wozniak H., Friedli D., Martinet L., Martin P., 
Bratschi P., 1995, A\&ASS, 111, 115  

\end{thebibliography}
\end{document}